\documentclass[prb,preprint,longbibliography]{revtex4-1} 

\usepackage{amsmath}  
\usepackage{amsfonts} 
\usepackage{graphicx} 
\usepackage{hyperref}
\usepackage{xcolor}
\usepackage[caption = false]{subfig}

\begin{document}


\title{On the physical (im)possibility of lightsabers}

\author{Fran\c{c}ois Fillion-Gourdeau}
\email{francois.fillion@emt.inrs.ca} 
\affiliation{Institute for Quantum Computing, University of Waterloo, Waterloo, Ontario, Canada}

\author{Jean-S\'{e}bastien Gagnon}
\email{gagnon01@fas.harvard.edu}
\affiliation{Department of Earth and Planetary Sciences, Harvard University, Cambridge, Massachusetts, USA}
\affiliation{Natural Sciences Department, Castleton University, Castleton, Vermont, USA}


\date{\today}

\begin{abstract}
In this paper, we use a science fiction theme (i.e. the iconic lightsaber from the Star Wars universe) as a pedagogical tool to introduce aspects of nonlinear electrodynamics due to the quantum vacuum to an audience with an undergraduate physics background.  In particular, we focus on one major problem with lightsabers that is commonly invoked as an argument to dismiss them as unrealistic: light blades are not solid and thus cannot be used in a duel as normal swords would.  Using techniques coming from ultra intense laser science, we show that for high enough laser intensities, two lightsaber blades can ``feel'' solid to each other.  We argue that this aspect of  lightsabers is not  impossible  due  to
limitations of the laws of physics, but is very implausible due to the high intensities and
energy needed for their operation.

\end{abstract}

\maketitle 

\section{Introduction} 
\label{sec:Introduction}

``This is the weapon of a Jedi Knight. Not as clumsy or random as a blaster. An elegant weapon... for a more civilized age.''  The above quote by Obi-Wan Kenobi introduced the now iconic lightsaber to the world in the first Star Wars movie of 1977.  It immediately struck the imagination of many, spurring a wave of interest in fans, who started to build replicas and even choreograph lightsaber duels.

Out of all this excitement came the natural question ``Are lightsabers possible?''  According to the Star Wars movies, a lightsaber is very similar to a conventional sword in which the steel blade has been replaced with a blade made of light.  It can cut through steel, reflect laser shots coming from blasters, and be used in duels just like conventional swords.  Despite the high expectations of fans, there are serious problems with such a device\cite{IOP_Problems_with_lightsabers,Speculations_on_lightsabers}.  For instance, a finite light blade is implausible, since light cannot abruptly stop on its own after propagating for a certain length.  Another major problem is the apparent solidity of light blades when hitting each other, which seems contrary to our day-to-day experience with light.

Due to the above problems, the focus changed to inventing similar devices with more plausible designs and sharing many characteristics with lightsabers (e.g. finite blades, apparent solidity of the blades, melting power, etc).  A well-known example is the ``plasma saber'' design of Michio Kaku\cite{Kaku_2009,SciFiScience}, where the blade is made of a ceramic cylinder punctured with holes from which a high temperature plasma is projected and confined using magnetic fields.  

The plasma saber design is interesting on its own, but it is unlikely that its solid ceramic blade could retract into the hilt as depicted in the movie.  This is a minor point, but it could be unappealing for certain Star Wars fans.  For this reason, it is worthwhile to go back to the original lightsaber design, and see if there are any {\em fundamental} limitations in the laws of physics that would prevent their construction.

As mentioned previously, the first fundamental problem is the finite length of the light blade.  Thus to build a functional lightsaber, a mechanism for stopping light is necessary.  It is known that light  effectively  slows down when passing through matter, with the new speed of light given by $c/n$ (where $n$ is the index of refraction).  Using more sophisticated techniques, Lau and collaborators\cite{Lau_etal_1999,Liu_etal_2001,Lau_2001} showed that it is possible to stop light going through a cloud of cold atoms.  These experiments are a far cry from being able to build a real lightsaber, but they show that it is possible to manipulate light in a way akin to what is depicted in the Star Wars movies,  as speculated by Dolors on her website ``Cracking the nutshell''~\cite{Speculations_on_lightsabers}. 

The second fundamental problem is the apparent solidity of the blades when hitting each other.  Classically, light in vacuum does not interact with itself, implying that a light blade cannot be used to stop another light blade.  Thus some additional ingredients are required to have a lightsaber-like behavior where the two blades interact. This is possible when photons are propagating in media with some specific properties. For instance, when pairs of photons are shot into a specially prepared cloud of cold atoms, a medium-mediated interaction between the photons is created\cite{Firstenberg_etal_2013,Photonic_molecules}.  Owing to this interaction, the photons are pushing and pulling each other, a necessary condition to have light blades being able to hit one another.

The main issue with the above approach to the two fundamental problems of lightsabers is that it requires the presence of a (nonlinear) medium in which light has to propagate.  In the above cases, the medium is a gas of atoms maintained at a few nanokelvins, a feat difficult to achieve in the context of lightsabers, requiring the presence of some sort of refrigerated case around the lightsaber blade to contain the gas.  This would be highly impractical.  Thus a valid question to ask is ``Is there a way for light to have lightsaber behavior in vacuum?''.

 Concerning the obtention of a finite blade in vacuum, it has been shown by various groups\cite{Kitamura_etal_2010,Dehez_etal_2012,Panneton_etal_2015} that it is possible to obtain ``needles of light'' by tightly focusing an annular beam of radially polarized light.  In this configuration, intricate interference effects make for a diffraction pattern with an axially elongated shape where most of the electromagnetic energy is concentrated. As the name implies, these needles of light are of finite extent both longitudinally and transversally. Up to now, lengths of a few thousands wavelengths have been achieved\cite{Panneton_etal_2015} (about $1$ mm for visible wavelengths).  This is clearly insufficient to build the lightsabers portrayed in the Star Wars movies, but shows that a finite blade could in principle be achieved.  

 As for the apparent solidity of the blades in vacuum, quantum physics offers a possible solution.  \color{black}  As  mentioned \color{black} by the physicist Brian Cox in an interview with Neil deGrasse Tyson\cite{StarTalk}, contrary to popular belief light can actually interact with itself in vacuum.  This interaction (first predicted by Euler and Kochel in 1935\cite{Euler1935}) is due to quantum effects, as we discuss in Sect.~\ref{sec:Nonlinear_em}.   This light-by-light scattering has been directly measured at the Large Hadron Collider by the ATLAS collaboration\cite{ATLAS_2017_Light_by_light_scattering}.  The cross section for photon-photon scattering is maximal for photons with an energy equal to the rest mass energy of the electron $ mc^{2}$ (where $m$ is the mass of the electron and $c$ is the speed of light), corresponding to a frequency in the lower range of gamma rays\cite{Marklund_Shukla_2006}. Therefore, one could contemplate ``gamma ray sabers'' that would interact with each other when passing through one another. However, these considerations do not explain precisely the physical process responsible for a possible ``recoil'' effect occurring when the lightsabers are crossing, making for the apparent solidity of the blades.     

In addition, lightsabers are made of visible light, not gamma rays.  Thus to stay faithful to the spirit of the Star Wars movies, it is imperative to study the apparent solidity of light blades for wavelengths in the visible range.  In this paper, we study the effects of light-by-light scattering when two (faithful to the movies) lightsaber blades cross paths.  In particular, we investigate the force felt on one lightsaber hilt due to the scattered light coming from the interaction with the other lightsaber.  This constitutes our main science goal.  But our (hidden) pedagogical goal here is to demonstrate that examples coming from science-fiction and fantasy can be used as pedagogical tools to introduce advanced physics to undergraduates (in the present case, nonlinear effects in electrodynamics due to the quantum vacuum).  This idea of using science-fiction themes in the classroom has already been exploited with great success by Munz and collaborators\cite{Munz_etal_2009,Smith_2014} and others \cite{Alemi_etal_2015}, who modelled a zombie attack based on biological assumptions coming from popular movies. Similarly, the merits of using science-fiction movies for teaching the intricate physics of wormholes in general relativity to undergraduate students have been assessed by Thorne and his collaborators \cite{doi:10.1119/1.4916949,doi:10.1119/1.15620}.  The University of Leicester pushes this idea further, and encourages their senior physics students to choose a ``special topic'' (often coming from science-fiction or fantasy), analyze it quantitatively and write a paper about their findings\cite{Special_topics}.  In the same spirit, we argue that allowing science-fiction and fantasy examples to be seriously analyzed and discussed in the classroom opens up all new vistas for the teaching of physics, and this paper is an example of this.

And above all else, who can really resist the lure of the dark side...

\section{Theoretical background}
\label{sec:Theorectical_background}

To formulate our lightsaber example in a precise way, it is essential to understand the origin of light-by-light scattering in vacuum. To reach this goal, we first review the salient features of classical electrodynamics, and then incorporate nonlinear effects coming from the quantum vacuum.

\subsection{Classical electrodynamics}
\label{sec:Classical_em}

In the absence of free charges and currents, the differential Maxwell's equations take the form\cite{Griffiths_2013}:
\begin{eqnarray}
\label{eq:Maxwell_1}
\nabla\cdot \mathbf{E}(\mathbf{r},t) & = & -\frac{1}{\epsilon_{0}}\nabla\cdot \mathbf{P}(\mathbf{r},t), \\
\label{eq:Maxwell_2}
\nabla\cdot \mathbf{B}(\mathbf{r},t) & = & 0, \\
\label{eq:Maxwell_3}
\nabla \times \mathbf{E}(\mathbf{r},t) & = & -\frac{\partial \mathbf{B}(\mathbf{r},t)}{\partial t}, \\
\label{eq:Maxwell_4}
\nabla \times \mathbf{B}(\mathbf{r},t) & = & \frac{1}{c^{2}}\frac{\partial \mathbf{E}(\mathbf{r},t)}{\partial t} + \mu_{0}\left(\nabla\times \mathbf{M}(\mathbf{r},t) + \frac{\partial \mathbf{P}(\mathbf{r},t)}{\partial t}\right),
\end{eqnarray}
where $\bf{E}$ and $\bf{B}$ are the electric and magnetic vector fields, $\bf{P}$ and $\bf{M}$ are the polarization and magnetization vector fields, $\epsilon_{0}$ and $\mu_{0}$ are the permittivity and permeability of free space, respectively, and $c$ is the speed of light. These equations yield a unique solution when they are complemented by some initial condition ($\mathbf{E}(t_{\mathrm{init}})$ and $\mathbf{B}(t_{\mathrm{init}})$, where $t_{\mathrm{init}}$ is the initial time).  This solution fully describes the dynamical and spatial behavior of a classical electromagnetic field. 

Taking the curl of Eqs.~(\ref{eq:Maxwell_3}) and (\ref{eq:Maxwell_4}) and using vector identities, we obtain the following wave equations\cite{Griffiths_2013}:
\begin{eqnarray}
\label{eq:Wave_equation_1}
\left(\frac{1}{c^{2}}\frac{\partial^{2} }{\partial t^{2}} - \nabla^{2}\right)\mathbf{E}(\mathbf{r},t) & = & \mathbf{S}_{1}(\mathbf{r},t), \\
\label{eq:Wave_equation_2}
\left(\frac{1}{c^{2}}\frac{\partial^{2} }{\partial t^{2}} - \nabla^{2}\right)\mathbf{B}(\mathbf{r},t) & = & \mathbf{S}_{2}(\mathbf{r},t),
\end{eqnarray}
with source terms given by:
\begin{eqnarray}
\label{eq:Source_term_1}
\mathbf{S}_{1}(\mathbf{r},t) & = & \mu_{0}\left(-\frac{\partial [\nabla\times\mathbf{M}(\mathbf{r},t)]}{\partial t} - \frac{\partial^{2} \mathbf{P}(\mathbf{r},t)}{\partial t^{2}} + c^{2}\nabla[\nabla\cdot\mathbf{P}(\mathbf{r},t)]\right), \\
\label{eq:Source_term_2}
\mathbf{S}_{2}(\mathbf{r},t) & = & \mu_{0}\left(\frac{\partial[\nabla\times\mathbf{P}(\mathbf{r},t)]}{\partial t} - \nabla^{2}\mathbf{M}(\mathbf{r},t) + \nabla[\nabla\cdot\mathbf{M}(\mathbf{r},t)] \right).
\end{eqnarray}
Equations~(\ref{eq:Wave_equation_1})-(\ref{eq:Wave_equation_2}) describe propagating electromagnetic waves sourced by the polarization $\mathbf{P}$ and the magnetization $\mathbf{M}$.  For example, when an incident electromagnetic wave enters matter, it can deform orbitals and reorient magnetic dipoles.  These local changes then feedback into Eqs.~(\ref{eq:Wave_equation_1})-(\ref{eq:Wave_equation_2}) through $\mathbf{P}$ and $\mathbf{M}$, and may affect the outgoing wave.

Classically and in the absence of matter, both $\mathbf{P}$ and $\mathbf{M}$ are zero.  In this case, the wave equations~(\ref{eq:Wave_equation_1})-(\ref{eq:Wave_equation_2}) are linear in the  fields $\mathbf{E}$ and $\mathbf{B}$ (i.e. if $\mathbf{E}_{1}$ and $\mathbf{E}_{2}$ are both solutions to Eq.~(\ref{eq:Wave_equation_1}), then $c_{1}\mathbf{E}_{1} +c_{2}\mathbf{E}_{2}$ is also a solution).  The linearity property of the equations underlies the principle of superposition of waves.  This implies that, when two incoming electromagnetic waves ``collide'' with each other, their amplitudes add up for a brief instant during the collision, and then leave the collision area unscathed.  Thus classically and in vacuum, two light beams cannot interact with each other due to the linearity property of Maxwell's equations.  This is the main reason why the apparent solidity of lightsabers is generally thought to be physically impossible.

\subsection{Nonlinear electrodynamics and the quantum vacuum}
\label{sec:Nonlinear_em}

The linearity property of Maxwell's equations still holds in the presence of many types of media.  In particular, electromagnetic waves propagating through some material deforms the orbitals of its constituent atoms, thus inducing an atomic polarization.  When the electric field is weak, these deformations are small, and the polarization is proportional to the electric field (a similar argument can be made for the magnetization).  In such a case, the sources $\mathbf{S}_{1}$ and $\mathbf{S}_{2}$ are  linear in the field, implying that Maxwell's equation in (linear) media obey the superposition principle.  However, when the electric field is strong, the orbital deformations are large, and the polarization is no longer linear in the electric field. The physical origin of this nonlinearity can be traced back to the details of the interaction binding the electron to the atomic nucleus \cite{Boyd_2003}. Then, the polarization is usually written as a power series in the electric field\cite{Boyd_2003}:
\begin{eqnarray}
\label{eq:Taylor_expansion_polarization}
|\mathbf{P}| & = & \chi^{(1)}|\mathbf{E}| + \chi^{(2)}|\mathbf{E}|^{2} + \chi^{(3)}|\mathbf{E}|^{3} + \dots
\end{eqnarray}
where $\chi^{(i)}$ is $i^{\rm th}$ order susceptibility of the material (these coefficients encode the microscopic physics involved in the deformation of orbitals, and depend on the nature of the material or the type of matter subjected to the radiation).  When the electric field is strong, higher order terms are non-negligible compared to the first one, and the medium is considered to be nonlinear.

Consequently, as an electromagnetic wave propagates into matter, it modifies the local properties (e.g. orbital shapes) of the medium and thus, affects its own propagation.  If the intensity of the
wave is sufficiently large, a second electromagnetic wave passing through the medium would also be affected by the changes in the local properties induced by the first wave.  Thus even if the two electromagnetic waves do not interact directly, they can still interact indirectly through the presence of the medium.  Mathematically, this medium-mediated light-by-light effective interaction is represented by the nonlinear terms in the sources $\mathbf{S}_{1}$ and $\mathbf{S}_{2}$ (see Eq.~(\ref{eq:Wave_equation_1})-(\ref{eq:Wave_equation_2})).

\begin{figure}[h]
\centering
\includegraphics[height=4in]{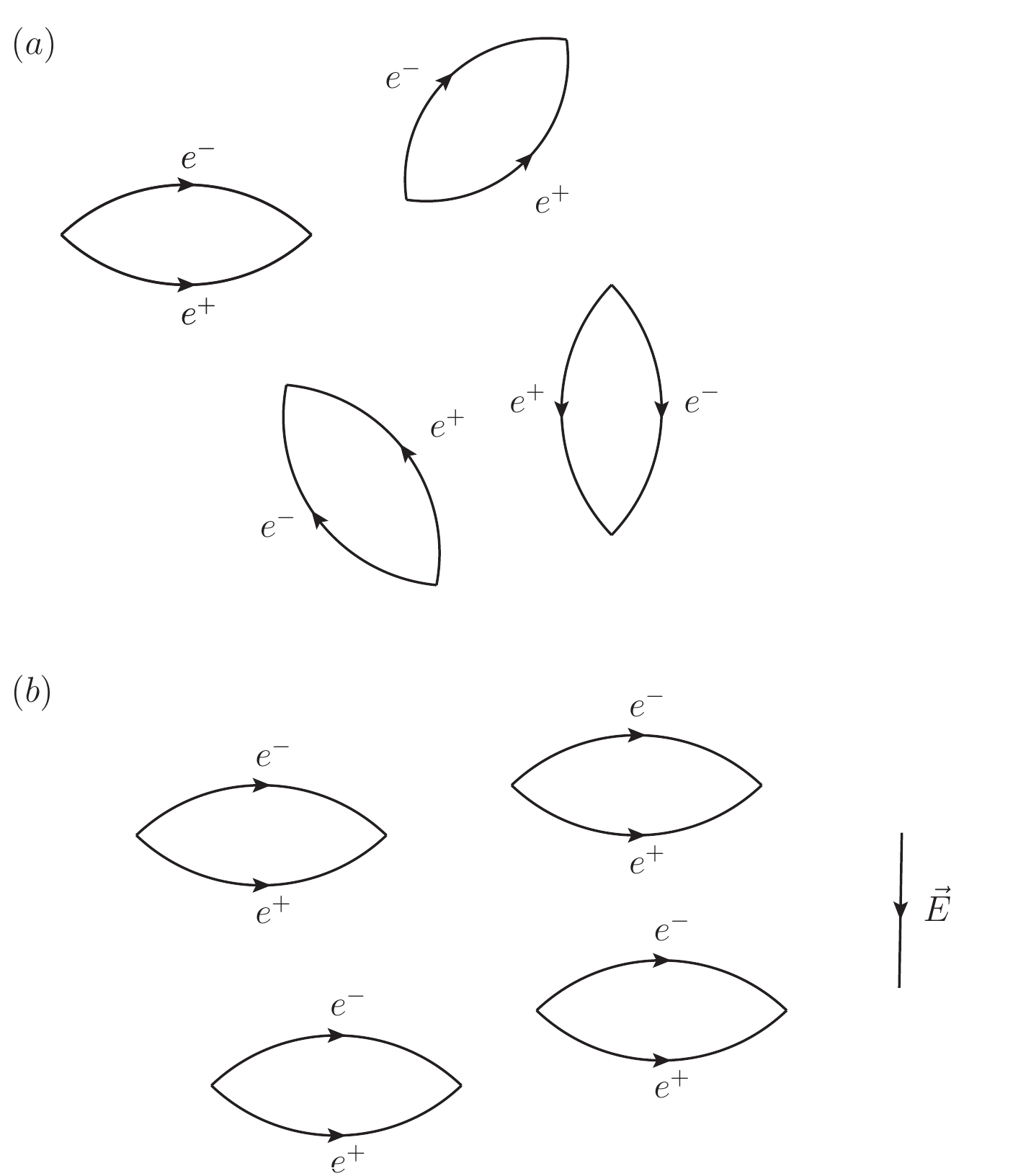}
\caption{(a) Illustration of virtual electron-positron pairs popping out of the quantum vacuum. (b) Polarized electron-positron pairs in the presence of an electric field}
\label{fig:Virtual_pairs}
\end{figure}

The vacuum in classical physics is absolutely empty.  We thus conclude from the above argument that light-by-light scattering (and consequently lightsabers) are not possible in a classical vacuum.  However, the conclusion is different when quantum effects are taken into account.  From progress in quantum field theory, it is generally accepted that the quantum vacuum is filled with virtual particles\cite{Peskin_Schroeder_1995}.  Those virtual particles are a consequence of the relation $E = mc^{2}$ coming from special relativity and the Heisenberg uncertainty principle $\Delta t/ \Delta E \leq \hbar$.  As long as charge conservation is not violated, energy can be ``borrowed'' from the vacuum to create virtual particles (such as electron-positron pairs) and returned to the vacuum in a time short enough so as to not contradict the Heisenberg uncertainty principle.  The ultra-short existence of those virtual particles prevents them from being directly detected, although their indirect effect have been measured very precisely (e.g. Lamb shift\cite{Peskin_Schroeder_1995}).  See Fig.~\ref{fig:Virtual_pairs} (a) for an illustration of virtual electron-positron pairs popping out of the vacuum.  

These virtual electron-positron pairs can be viewed as small electric dipoles.  The passage of an electromagnetic wave can thus polarize the quantum vacuum.  The situation is very similar to electrodynamics in matter, with the quantum vacuum playing the role of the nonlinear material  and virtual electron-positron pairs playing the role of  atomic orbitals.  It is thus possible to have light-by-light scattering mediated by virtual pairs in the vacuum.  This effect was first predicted in 1935 by Euler and Kochel \cite{Euler1935}, and further studied in 1936 by Heisenberg and Euler in a seminal paper that gave birth to quantum electrodynamics\cite{Heisenberg_Euler_1936}.  A calculation (beyond the scope of this paper) based on the Euler-Heisenberg theory shows that the polarization and magnetization due to quantum vacuum effects are\cite{Marklund_Shukla_2006}:
\begin{equation}
\label{eq:pol}
\mathbf{P}(\mathbf{r},t)  =  a \biggl\{ 2\left[\mathbf{E}^{2}(\mathbf{r},t) - c^{2}\mathbf{B}^{2}(\mathbf{r},t)   \right] \mathbf{E}(\mathbf{r},t)+ 7c^{2} \left[\mathbf{E}(\mathbf{r},t) \cdot \mathbf{B}(\mathbf{r},t) \right]\mathbf{B}(\mathbf{r},t) \biggr\} + O(a^{2}),
\end{equation}
\begin{equation}
\label{eq:mag}
\mathbf{M}(\mathbf{r},t) = a \biggl\{ 2c^{2}\left[c^{2}\mathbf{B}^{2}(\mathbf{r},t) - \mathbf{E}^{2}(\mathbf{r},t)   \right] \mathbf{B}(\mathbf{r},t)+ 7c^{2} \left[\mathbf{E}(\mathbf{r},t) \cdot \mathbf{B}(\mathbf{r},t) \right] \mathbf{E}(\mathbf{r},t) \biggr\} + O(a^{2}),
\end{equation}
where $a \equiv \frac{4\alpha^{2}\hbar^{3}\epsilon_{0}^{2}}{45m_{e}^{4}c^{5}}$, $\alpha \equiv \frac{e^{2}}{4\pi\epsilon_{0}\hbar c}$ is the fine structure constant, $\hbar$ is the reduced Planck constant, $e$ is the electron charge, $m_{e}$ is the electron mass, and $O(a^{2})$ corresponds to higher order terms in the electric and magnetic fields.  Note that Eqs.~(\ref{eq:pol})-(\ref{eq:mag}) are valid in the limit of low photon energy $\hbar\omega \ll m_{e}c^{2}$ and relatively small electric fields $|\mathbf{E}| \ll E_{\rm crit} \equiv \frac{2m_{e}c^{2}}{e\lambda_{c}} \approx 1.3 \times 10^{18}$~V/m, where $\lambda_{c}$ is the Compton wavelength of the electron\footnote{The critical electric field (also called the Schwinger field) can be heuristically obtained by equating the work necessary to separate a virtual electron-positron pair by a Compton wavelength ($eE_{\rm crit}\lambda_{c}$) to the energy required to produce an electron-positron pair ($2m_{e}c^{2}$).}.  When the field strength approaches the Schwinger field ($|\mathbf{E}| \lesssim E_{\rm crit}$), higher order terms in the fields become important\cite{dunne2005heisenberg}.  For higher photon energies or even higher electric fields $|\mathbf{E}| \gtrsim E_{\rm crit}$, real electron-positron pairs become important, and the full theory of quantum electrodynamics (QED) is necessary to perform calculations. The correspondence between QED and the Euler-Heisenberg theory was first put on firm ground by Karplus and Neuman \cite{PhysRev.80.380}, where the cross-section for photon-photon scattering in the Euler-Heisenberg theory was derived from quantum electrodynamics and modern Feynman diagrams techniques. This calculation demonstrated that the Euler-Heisenberg theory is actually the quintessential example of an effective low energy theory, whereby some degrees of freedom (here, the electrons and positrons) are ``integrated out'' to obtain an approximation of the full theory (here QED) in the low energy regime. Following this procedure, the complicated light-by-light scattering diagram is replaced by a simpler effective vertex that encodes the low energy scattering (see Fig. \ref{fig:interaction_vertex}). The polarization and magnetization in Eqs. \eqref{eq:pol} and \eqref{eq:mag} can be obtained from the latter.

\newsavebox{\tempbox}
\begin{figure}
\begin{center}
\hspace*{6.0em}
\sbox{\tempbox}{\includegraphics{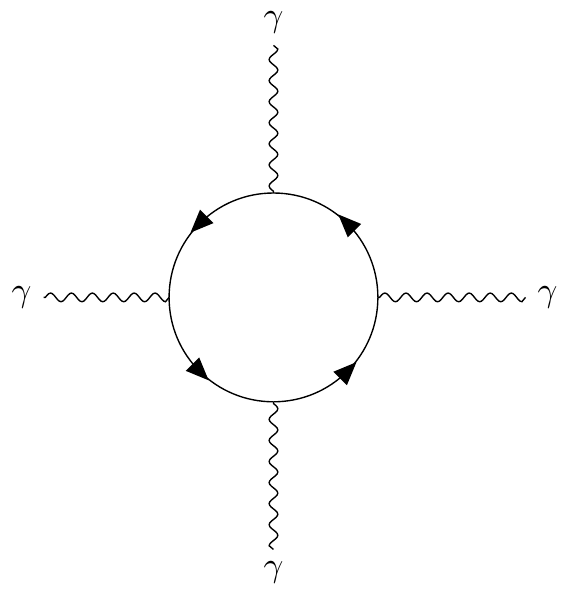}}
\subfloat[]{\usebox{\tempbox}} \hspace*{-15.0em}
\subfloat[]{\vbox to \ht\tempbox{\vfil \includegraphics{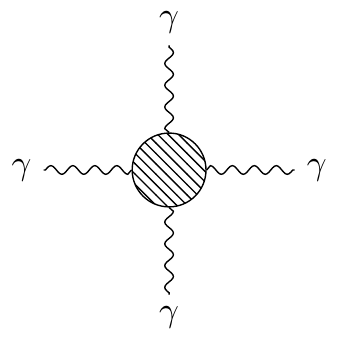} \vfil}}
\caption{Light-by-light scattering vertices in (a) QED and (b) Euler-Heisenberg effective theory.  Wiggly lines represent photons, full lines represent electrons and the blob represents the effective interaction. The mathematical expression of the QED vertex reduces to the effective Euler-Heisenberg expression in the limit where the photon energies obey $\hbar \omega \ll mc^{2}$. The QED theory clarifies that the interaction is mediated by the exchange of virtual electron-positron pairs. }
\label{fig:interaction_vertex}
\end{center}
\end{figure}

Note also that the polarization and magnetization effects due to the quantum vacuum are proportional to $a\approx 2.6\times 10^{-52}$ F$\cdot$m/J and are thus very small.  This makes these effects very hard to observe, unless the electric field strengths involved are close to $E_{\rm crit}$. For many decades after photon-photon interaction was proposed by Euler and Kochel, it was thought that these high fields could not be reached in laboratories. However, in the last few decades, there has been many technical developments in laser science that allow for unprecedented light intensity levels. The chirped-pulse-amplification technique \cite{STRICKLAND1985447}, for which the 2018 Nobel prize in physics was awarded, has been instrumental in this quest for ultrahigh intensity laser light. In current laboratories using petawatt class lasers \cite{danson2015}, it is now possible to reach an intensity of $\approx 10^{22}$ W/cm$^2$ (corresponding to a field strength of $E \approx 1.9 \times 10^{14}$ V/m) \cite{Bahk:04}.  Future facilities such as the Extreme Light Infrastructure (ELI) \cite{ELI}, the Apollon laser \cite{papadopoulos2016},  the Exawatt Center for Extreme Light Studies (XCELS) \cite{XCELS} and the High Power laser Energy Research (HiPER) \cite{HiPER} will be multi-petawatts infrastructures aiming at even higher intensities, on the order of $10^{23} - 10^{25}$ W/cm$^2$.  Motivated by these technological advances, theorists have proposed many schemes to detect light-by-light scattering using laser fields, taking advantage of analogies of the vacuum with nonlinear optics \cite{0034-4885-76-1-016401,king_heinzl_2016}. Some examples include the four-wave mixing process in the crossing of two \cite{PhysRevLett.97.083603,1367-2630-14-10-103002} or three laser beams \cite{PhysRevLett.96.083602,PhysRevA.74.043821}, vacuum processes in tightly focused laser fields \cite{PhysRevLett.107.073602,PhysRevA.91.031801}, birefringence \cite{HEINZL2006318,PhysRevLett.119.250403,1402-4896-91-2-023010,PhysRevD.92.071301,Dinu:2013gaa,Dinu:2014tsa} and self-focusing \cite{PhysRevA.62.043817}. Experimental attempts were performed by a french group using the four-wave-mixing process  \cite{moulin2000} and by the PVLAS collaboration using vacuum birefringence \cite{DellaValle2016}, but these investigations have been inconclusive so far.

Outside of laser science however, there is compelling evidence that light-by-light scattering is a real physical phenomenon. As alluded in the introduction, the ATLAS collaboration at the Large Hadron Collider have made a measurement of this phenomenon using heavy-ion collisions \cite{ATLAS_2017_Light_by_light_scattering}, albeit the process studied considers virtual photons (so-called quasi-real photons) in the initial state. Very high fields can be found in astrophysical systems, such as neutron stars. Hints towards the existence of vacuum birefringence were found by looking at the polarization of the x-ray radiation emitted from such a system \cite{mignani2016evidence}. Finally, indirect evidence was observed in the anomalous magnetic moment of the muon \cite{PhysRevD.73.072003}, whereby higher order corrections in the theoretical calculation demand light-by-light scattering Feynman diagrams. 

Note that in this paper, we focus on a specific QED process where light-by-light scattering is mediated by virtual electron-positron pairs. According to the Standard Model of particle physics, other particles can induce photon-photon interactions, such as pions or Higgs bosons.  However, the interaction cross-sections of these possible physical process are negligible in the regime studied in this article. 




\section{Electromagnetic fields generated by crossing laser beams}
\label{sec:EM_crossing_beams}

As argued in Sect.~\ref{sec:Nonlinear_em}, it is possible for two beams of light to interact with each other in vacuum.  In principle, this could give lightsaber wielders the sensation of apparent solidity of the blades when two lightsabers come into contact.  Equipped with Eqs.~(\ref{eq:Wave_equation_1})-(\ref{eq:Source_term_2}) and (\ref{eq:pol})-(\ref{eq:mag}), we are now in a position to formulate this situation more precisely.

\begin{figure}[ht]
\centering
\includegraphics[height=5in]{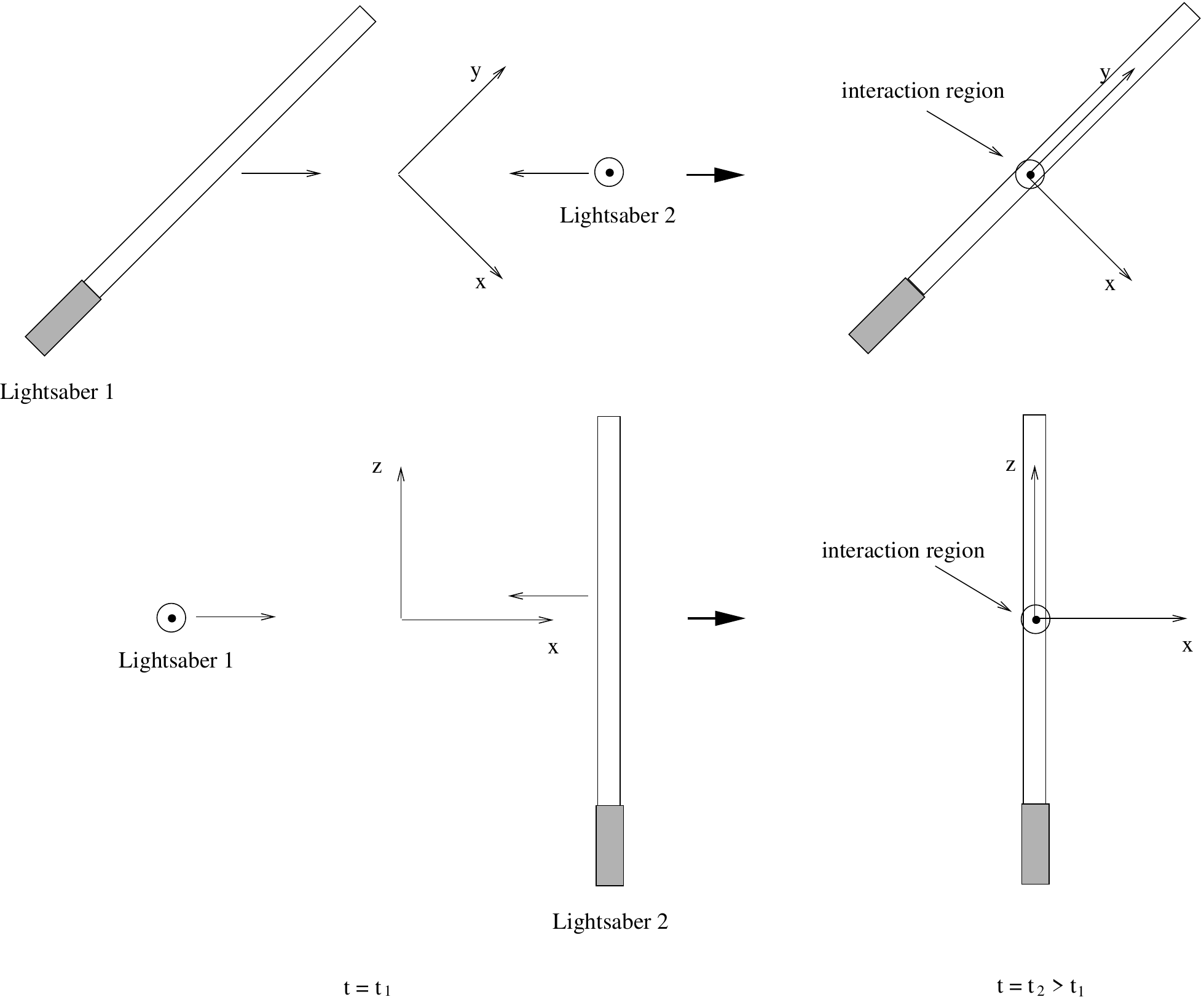}
\caption{Lightsaber geometry considered in this paper.  At time $t_{1}$ (left column), two lightsabers (labeled 1 and 2) approach each other (top and bottom rows correspond to different viewpoints).  At a later time $t_{2} > t_{1}$ (right column), the two lightsabers come into contact, defining a small interaction region where the two beams cross.}
\label{fig:Crossing_beams_geometry}
\end{figure}

For definiteness, consider the lightsaber geometry shown in Fig.~\ref{fig:Crossing_beams_geometry}.  When the two lightsaber beams come into contact, the overlap between them defines a small interaction region.  From this, the lightsaber interaction proceeds in the following way:
\begin{description}
    \item[Step 1] The lightsabers emit light at the hilt,  similar to an ultrahigh intensity laser.

    \item[Step 2] The light from each lighsaber propagates in vacuum and reaches the interaction region.

    \item[Step 3] Owing to photon-photon interactions and the structure of the vacuum, the incoming light is ``reflected'' in the interaction region (an effect called the vacuum mirror\cite{1367-2630-15-8-083002,1367-2630-17-4-043060}).  Thus in the interaction region, nonlinear effects due to the quantum vacuum generate new radiation in all directions.
 
    \item[Step 4] Some of the radiation emitted from one lightsaber is either reflected back or toward the other lightsaber.  This reflected radiation propagates from the interaction region to the hilt of either lightsabers and henceforth, produces radiation pressure at the hilt. 
\end{description}
If the recoil effect due to the radiation pressure of the reflected radiation is large enough, this can then be perceived as apparent solidity of the blades by the wielder.  Thus our goal is to compute the force at the hilt of one lightsaber owing to the reflected light coming from the interaction region.  To evaluate this, we need a model for the light emitted by each lightsaber, and a technique to compute the radiation generated in the interaction region due to the quantum vacuum.  We discuss each of these ingredients in the following.

\subsection{Lightsaber beam model}

To describe theoretically the light emitted from the lightsabers, we make the following assumptions:

\begin{description}
    \item[A1] The light emitted from each lightsaber is a square beam (of size $b$) that can be described using plane waves.  Mathematically, it corresponds to multiplying plane waves with a cutoff function $F(\mathbf{r})$.  It should be emphasized that this procedure simplifies the calculation significantly but cannot be taken as an accurate model for light propagation. In particular, it does not obey Maxwell's equations  at the edges of the lightsaber beam.  Nevertheless, it captures the main physical effects involved in lightsabers interaction.
    
    \item[A2] The interaction region is located at a distance $d \gg b$ from both hilts, and the collision between the blades occurs at 90$^{\circ}$. This is by no means representative of a real fight where the angle can vary in the interval $[0^{\circ},180^{\circ}]$ and the blades can overlap at any distance from the hilts, but it simplifies the calculation and is enough to obtain an order-of-magnitude estimate of the recoil effect.
    
    \item[A3] The polarization of the light emitted by each lightsaber is in the same direction (chosen to be $x$ in this case, see Fig.~\ref{fig:Crossing_beams_geometry}).  This assumption simplifies the calculation and is sufficient to get an order-of-magnitude estimate of the recoil effect, but is by no means required.
    
    \item[A4] A perfect vacuum is considered, preventing nonlinear effects due to light propagation in air. The latter would distort the propagating light through many nonlinear effects such as self-channeling, filamentation and spectral modulation \cite{Braun:95}. 
    
\end{description}


From the above considerations and following the geometry shown in Fig.~\ref{fig:Crossing_beams_geometry}, we write the light emitted at the hilt of each lightsaber as two plane waves propagating in the $\hat{\mathbf{y}}$ and $\hat{\mathbf{z}}$ direction, respectively ($\hat{\mathbf{x}},\hat{\mathbf{y}}$ and $\hat{\mathbf{z}}$ are unit vectors). The electromagnetic field of the plane waves is given by:
\begin{align}
\label{eq:E1}
\mathbf{E}_{1}(\mathbf{r},t) &= F_{1}(\mathbf{r}) \frac{E_{1}}{2} \left( e^{i(k_{1}y - \omega_{1} t)} + e^{-i(k_{1}y - \omega_{1} t)} \right)\hat{\mathbf{x}}, \\
\label{eq:B1}
\mathbf{B}_{1}(\mathbf{r},t) &= -F_{1}(\mathbf{r}) \frac{E_{1}}{2c} \left( e^{i(k_{1}y - \omega_{1} t)} + e^{-i(k_{1}y - \omega_{1} t)} \right)\hat{\mathbf{z}}, \\
\label{eq:E2}
\mathbf{E}_{2}(\mathbf{r},t) &= F_{2}(\mathbf{r}) \frac{E_{2}}{2} \left( e^{i(k_{2}z - \omega_{2} t)} + e^{-i(k_{2}z - \omega_{2} t)} \right)\hat{\mathbf{x}},
\\
\label{eq:B2}
\mathbf{B}_{2}(\mathbf{r},t) &= F_{2}(\mathbf{r}) \frac{E_{2}}{2c} \left( e^{i(k_{2}z - \omega_{2} t)} + e^{-i(k_{2}z - \omega_{2} t)} \right)\hat{\mathbf{y}},
\end{align}
%
%
%
%
where $E_{1,2}$ are the electric field strengths of each beam, $\omega_{1,2}$ are the angular frequencies of the two beams, and $k_{1,2}$ are their wave vector magnitudes. The cutoff functions are given by:
\begin{align}
    F_{1}(\mathbf{r}) & = \mathrm{rect}\left(\frac{x}{b}\right)
    \mathrm{rect}\left(\frac{z}{b}\right), \\
     F_{2}(\mathbf{r}) & = 
     \mathrm{rect}\left(\frac{x}{b}\right)
     \mathrm{rect}\left(\frac{y}{b}\right),
\end{align}
where $b$ is the size of the square beams and where the rectangular function is defined as:
\begin{align}
    \mathrm{rect}\left(\frac{\xi}{L}\right) := \theta\left(\xi + \frac{L}{2}\right) - \theta\left(\xi - \frac{L}{2}\right)  :=
    \begin{cases}
    1 & \mbox{if} \; |\xi| \leq \frac{L}{2} \\
    0 & \mbox{if} \; |\xi| > \frac{L}{2}
    \end{cases},
\end{align}
where $\theta(x)$ is the Heaviside step function.

\subsection{Radiation generated by the interaction region}

Once the model of the incident beam field is known, it is possible to compute the generated radiation coming from the interaction region. Note that outside the interaction region, the effect of the vacuum on the propagation of plane waves is negligible because the polarization and magnetization are zero there (this can be seen immediately by substituting Eqs.~(\ref{eq:E1})-(\ref{eq:B2}) into Eqs.~(\ref{eq:pol})-(\ref{eq:mag})). This is an important characteristic of plane waves: because of Lorentz invariance,  they cannot polarize the vacuum \cite{Schwinger:1951nm}.

To obtain the generated radiation coming from the interaction region, we solve the wave equations~(\ref{eq:Wave_equation_1})-(\ref{eq:Wave_equation_2}) by a linearization procedure, taking into account the incoming beams from the two lightsabers.  To do so, we first decompose the total electric and magnetic fields in two contributions:
\begin{eqnarray}
\label{eq:Total_E}
\mathbf{E}(\mathbf{r},t) &=&  \mathbf{E}_{\rm inc}(\mathbf{r},t) + \tilde{\mathbf{E}}(\mathbf{r},t), \\
\label{eq:Total_B}
\mathbf{B}(\mathbf{r},t) &=&  \mathbf{B}_{\rm inc}(\mathbf{r},t) + \tilde{\mathbf{B}}(\mathbf{r},t), 
\end{eqnarray}
where $\mathbf{E}_{\rm inc} = \mathbf{E}_{1} + \mathbf{E}_{2}$ and $\mathbf{B}_{\rm inc} = \mathbf{B}_{1} + \mathbf{B}_{2}$ are the incoming external fields coming from the lightsabers, while $\tilde{\mathbf{E}},\tilde{\mathbf{B}}$ are the fields generated by the nonlinear interaction.  From assumptions A1 and A4, we can write that the incoming fields (approximately) obey the homogeneous wave equations:
\begin{eqnarray}
\label{eq:Wave_equation_1_Ext}
\left(\frac{1}{c^{2}}\frac{\partial^{2} }{\partial t^{2}} - \nabla^{2}\right)\mathbf{E}_{\rm inc}(\mathbf{r},t) & \approx & 0, \\
\label{eq:Wave_equation_2_Bext}
\left(\frac{1}{c^{2}}\frac{\partial^{2} }{\partial t^{2}} - \nabla^{2}\right)\mathbf{B}_{\rm inc}(\mathbf{r},t) & \approx & 0.
\end{eqnarray}
This implies that the incoming fields from the two lightsabers are propagating freely while the effect of the nonlinear interaction is relegated to the fields $\tilde{\mathbf{E}},\tilde{\mathbf{B}}$, as shown below.

From our discussion of light-light interaction in Sect.~\ref{sec:Nonlinear_em}, we expect the generated fields to be much smaller than the fields coming from the lightsabers ($|\tilde{\mathbf{E}}| \ll |\mathbf{E}_{\rm inc}|$, $|\tilde{\mathbf{B}}| \ll |\mathbf{B}_{\rm inc}|$).  Substituting Eqs.~(\ref{eq:Total_E})-(\ref{eq:Total_B}) into Eqs.~(\ref{eq:Wave_equation_1})-(\ref{eq:Wave_equation_2}), using Eqs.~(\ref{eq:Wave_equation_1_Ext})-(\ref{eq:Wave_equation_2_Bext}) and keeping only the large contributions due to the incoming fields in the source terms $\mathbf{S}_{1}$ and $\mathbf{S}_{2}$, we obtain:
\begin{equation}
\label{eq:Wave_equation_1_Egenerated}
\left(\frac{1}{c^{2}}\frac{\partial^{2} }{\partial t^{2}} - \nabla^{2}\right)\tilde{\mathbf{E}}(\mathbf{r},t) = \mu_{0}\left(-\frac{\partial [\nabla\times\mathbf{M}_{\rm inc}(\mathbf{r},t)]}{\partial t} - \frac{\partial^{2} \mathbf{P}_{\rm inc}(\mathbf{r},t)}{\partial t^{2}} + c^{2}\nabla[\nabla\cdot\mathbf{P}_{\rm inc}(\mathbf{r},t)]\right),
\end{equation}
\begin{equation}
\label{eq:Wave_equation_2_Bgenerated}
\left(\frac{1}{c^{2}}\frac{\partial^{2} }{\partial t^{2}} - \nabla^{2}\right)\tilde{\mathbf{B}}(\mathbf{r},t) = \mu_{0}\left(\frac{\partial[\nabla\times\mathbf{P}_{\rm inc}(\mathbf{r},t)]}{\partial t} - \nabla^{2}\mathbf{M}_{\rm inc}(\mathbf{r},t) + \nabla[\nabla\cdot\mathbf{M}_{\rm inc}(\mathbf{r},t)] \right),
\end{equation}
where $\mathbf{P}_{\rm inc}$ and $\mathbf{M}_{\rm inc}$ are given by Eqs.~(\ref{eq:pol})-(\ref{eq:mag}) with $\mathbf{E}\rightarrow \mathbf{E}_{\rm inc}$ and $\mathbf{B}\rightarrow \mathbf{B}_{\rm inc}$.  Equations~(\ref{eq:Wave_equation_1_Egenerated})-(\ref{eq:Wave_equation_2_Bgenerated}) give the evolution of the generated fields, sourced by the external lightsaber beams through the nonlinear polarization and magnetization induced by the quantum vacuum.  These equations have to be solved in order to find the generated fields outside of the interaction region.  A detailed solution of Eqs.~(\ref{eq:Wave_equation_1_Egenerated})-(\ref{eq:Wave_equation_2_Bgenerated}) can be found in Appendix~\ref{app:Solutions_generated_fields}.  For this section, we instead use physical arguments and dimensional analysis to justify this solution.  

Physically, solving Eqs.~(\ref{eq:Wave_equation_1_Egenerated})-(\ref{eq:Wave_equation_2_Bgenerated}) amounts to finding the fields $\tilde{\mathbf{E}}$ and $\tilde{\mathbf{B}}$ generated by the source on the RHS of Eqs.~(\ref{eq:Wave_equation_1_Egenerated})-(\ref{eq:Wave_equation_2_Bgenerated}).  The source here is the cubic interaction region defined by the crossing laser beams (see Fig.~\ref{fig:Crossing_beams_geometry}), and we want the generated fields $\tilde{\mathbf{E}}$ and $\tilde{\mathbf{B}}$ at a point $\mathbf{r}$ located far away outside of the source (more precisely at the hilt of one of the lightsabers).

An important point here is that the source is time-dependent and oscillates, due to the fact that it is produced by the interaction of two plane waves that are themselves space-time dependent.  Thus each point in the source has varying  field induced polarization and magnetization, which generate spherical electromagnetic waves in all directions.  The total generated field at point $\mathbf{r}$ is thus the sum of all electromagnetic waves produced by all points in the source  modulated by the spatial dependence of the source itself.  For the electric field, this can be written as (the magnetic field is done in a similar way):
\begin{eqnarray}
\label{eq:Generated_field_approx_1}
\tilde{\mathbf{E}}(\mathbf{r},t) & \sim & \mu_{0}a E_{i}^{3}\int_{{\cal V}} d^{3}l\;  \frac{\omega^{2} e^{-i\omega t + ik|\mathbf{r} - \mathbf{l}|}}{|\mathbf{r}-\mathbf{l}|}\; e^{i\mathbf{k}'\cdot \mathbf{l}} \;\hat{\mathbf{x}},
\end{eqnarray}
where the prefactor $\mu_{0} a E_{i}^{3}$ can be read directly from Eqs.~(\ref{eq:Wave_equation_1_Egenerated})-(\ref{eq:Wave_equation_2_Bgenerated}), (\ref{eq:pol})-(\ref{eq:mag}) and (\ref{eq:E1})-(\ref{eq:B2}).  Note that the prefactor structure is reminiscent of the fact that the quantum vacuum induces a cubic nonlinearity.  The integral is over the volume ${\cal V}$ of the interaction region, i.e. the electromagnetic radiation generated at each point in the interaction region is added to obtain the full contribution.  The factor $e^{-i\omega t + ik|\mathbf{r} - \mathbf{l}|}/|\mathbf{r} - \mathbf{l}|$ represents a spherical wave of frequency $\omega$ and wavenumber $k$ (discussed in more details below)  propagating from a point~$\mathbf{l}$ in the source to the point $\mathbf{r}$, and the $\omega^{2}$ factor comes from substituting Eqs.~(\ref{eq:E1})-(\ref{eq:B2}) in Eqs.~(\ref{eq:pol})-(\ref{eq:mag}) (see below).  Since the source is produced by oscillating plane waves coming from the laser beams, it is reasonable to assume a sinusoidal spatial dependence for the source, represented by $e^{i\mathbf{k}'\cdot \mathbf{l}}$ in Eq.~(\ref{eq:Generated_field_approx_1}).   The polarization of the generated electric field is $\hat{\mathbf{x}}$, because the field induced polarization is in the same direction as the incoming electric field for plane waves.


Since we are interested in the generated electric field at a point far away from the source (see assumption A2), we can approximate $|\mathbf{r}-\mathbf{l}| \approx |\mathbf{r}| - \hat{\mathbf{r}}\cdot\mathbf{l}$, which gives:
\begin{eqnarray}
\tilde{\mathbf{E}}(\mathbf{r},t) & \sim & \mu_{0}a E_{i}^{3} \;\frac{\omega^{2} e^{-i\omega \left(t - \frac{|\mathbf{r}|}{c}\right)}}{|\mathbf{r}|} \int_{{\cal V}} d^{3}l\;  e^{ -ik(\hat{\mathbf{r}}\cdot\mathbf{l}) } \;e^{i\mathbf{k}'\cdot \mathbf{l}} \;\hat{\mathbf{x}}
\end{eqnarray}
where we neglected $O(b^{2}/d^{2})$ terms.  Note the appearance of the retarded time dependence in the first exponential, as expected from the finiteness of the speed of light.  Assuming we want the electric field at the hilt of saber 1, we can write $\mathbf{r} = -d\hat{\mathbf{y}}$, which gives:  
\begin{eqnarray}
\label{eq:Generated_E_field_approximate}
\tilde{\mathbf{E}}(-d\hat{\mathbf{y}},t) & \sim & \mu_{0}a E_{i}^{3} \;\frac{\omega^{2} e^{-i\omega \left(t - \frac{d}{c}\right)}}{d} \int_{-\frac{b}{2}}^{\frac{b}{2}}dx \int_{-\frac{b}{2}}^{\frac{b}{2}}dy \int_{-\frac{b}{2}}^{\frac{b}{2}}dz\;  e^{i(k y + \mathbf{k}'\cdot \mathbf{l})} \;\hat{\mathbf{x}} 
\end{eqnarray}
We expect the generated electromagnetic waves to be driven at the same frequency as the source, which itself depends on the frequency of the beams.  By substituting Eqs.~(\ref{eq:E1})-(\ref{eq:B2}) into Eqs.~(\ref{eq:pol})-(\ref{eq:mag}), a straightforward but tedious calculation shows that the frequency dependence of the source is $\omega^{2}e^{-i\omega t}$, where $\omega$ can take the following six values: $\omega_{1}$, $\omega_{2}$, $2\omega_{1} + \omega_{2}$, $\omega_{1} + 2\omega_{2}$, $2\omega_{1}-\omega_{2}$, $\omega_{1} - 2\omega_{2}$.  The above is a key difference between linear and nonlinear optics.  In linear optics, there is no mixing between frequencies, i.e. incoming photons entering the interaction region would keep their frequency in the absence of the nonlinear source terms.  The presence of nonlinear terms in the source allows multiple incoming photons to interact together and produce outgoing photons with different frequencies from the incoming ones.  Each of the above six frequencies correspond to a different nonlinear process and are well-known in nonlinear optics.  For instance, $\omega_{1}$ and $\omega_{2}$ correspond to parametric amplification (i.e. an amplification of the incoming field), $2\omega_{1} + \omega_{2}$ and $\omega_{1} + 2\omega_{2}$ are sum frequency mixing processes, and $2\omega_{1}-\omega_{2}$, $\omega_{1} - 2\omega_{2}$ are four-wave mixing processes.  Each of these processes have numerous applications in nonlinear optics, such as tuning the output frequency of a laser beam by letting it through a nonlinear crystal.  See Boyd\cite{Boyd_2003} for a discussion.

In nonlinear optics, the production of electromagnetic waves by shining light on a nonlinear crystal depends on the beam frequency, the crystal size and the spatial orientation of the beam with respect to the crystal lattice\cite{Boyd_2003}.  Typically, the production of electromagnetic waves is maximized in the direction where all phases in the second exponential in Eq.~(\ref{eq:Generated_E_field_approximate}) are zero  (i.e. when $ky + \mathbf{k}'\cdot\mathbf{l} = 0$).  When this so-called phase matching condition is fulfilled, the amplitude of the radiation generated from all points are in phase at the end of the interaction region, inducing the strongest possible signal. This phenomenon occurs in a specific direction where the photon momentum is conserved. In our case, the momentum $\mathbf{k}$ of the generated photons is conserved when it takes the value $\mathbf{k}_{1}$, $\mathbf{k}_{2}$, $2\mathbf{k}_{1} + \mathbf{k}_{2}$,$\mathbf{k}_{1} + 2\mathbf{k}_{2}$ ,  $2\mathbf{k}_{1}-\mathbf{k}_{2}$ and $\mathbf{k}_{1} - 2\mathbf{k}_{2}$ (these momentum values correspond to the frequencies enumerated previously).  As a consequence, depending on the geometry, only one driving frequency produces electromagnetic waves efficiently.  For the specific lightsaber geometry considered in Fig.~\ref{fig:Crossing_beams_geometry}, we have $\mathbf{k}_{1} = k_{1}\hat{\mathbf{y}}$ and $\mathbf{k}_{2} = k_{2}\hat{\mathbf{z}}$.  We thus conclude that it is not possible to satisfy the phase matching condition for any value of $\mathbf{k}$ and $\mathbf{k}'$ allowed by momentum conservation.  The best that can be achieved is a partial phase matching in the $z$-coordinate when $\mathbf{k} = \mathbf{k}' = \mathbf{k}_{1}$, giving $ky + \mathbf{k}'\cdot\mathbf{l} = 2k_{1}y$.  Thus the frequency mode $\omega_{1}$ should be more intense than the other modes.  
This is confirmed by a more detailed analysis (see Appendix~\ref{app:Solutions_generated_fields}), which shows that the beam frequency $\omega=\omega_{1}$ dominates the production of electromagnetic waves at the hilt of saber 1.  This indicates that a reflection of beam 1 toward hilt 1 is more efficient than a reflection of beam 2 toward hilt 1.  Taking this into account, the final result for the generated fields are (see Appendix~\ref{app:Solutions_generated_fields} for details):
\begin{eqnarray}
\label{eq:Solution_E_farfield_simplified_final}
\tilde{\mathbf{E}}(-d\hat{\mathbf{y}},t) & = & \frac{\mu_{0}}{2\pi}a  E_{1}E_{2}^{2}\; \omega_{1}^{2} \left(\frac{b^{2}}{d}\right)\frac{\sin(k_{1}b)}{k_{1}} \cos \left[\omega_{1}\left(\frac{d}{c} - t\right) \right] \hat{\mathbf{x}}.
\end{eqnarray}
The magnetic field at the hilt of saber 1 can be obtained in a similar way.  The result is:
\begin{eqnarray}
\label{eq:Solution_B_farfield_simplified_final}
\tilde{\mathbf{B}}(-d\hat{\mathbf{y}},t) & = & \frac{\mu_{0}}{2\pi}\frac{a}{c} E_{1}E_{2}^{2}\; \omega_{1}^{2}  \left(\frac{b^{2}}{d}\right)\frac{\sin(k_{1}b)}{k_{1}} \cos\left[ \omega_{1}\left(\frac{d}{c} - t\right)\right] \hat{\mathbf{z}}.
\end{eqnarray}
Note that the electric and magnetic fields at the hilt of saber 2 are obtained in the same way, and have similar expressions (with indices swapped $1 \leftrightarrow 2$, $\tilde{\mathbf{E}}$ in the $\hat{\mathbf{x}}$ direction and $\tilde{\mathbf{B}}$ in the $\hat{\mathbf{y}}$ direction).


\color{black}

\subsection{Radiation pressure at the hilts}

With expressions for the generated electric and magnetic fields when the two lightsaber beams overlap (c.f. Eqs.~(\ref{eq:Solution_E_farfield_simplified_final})-(\ref{eq:Solution_B_farfield_simplified_final})), we can compute the radiation pressure at the hilt of saber~1.  The radiation pressure $P$ can be computed using\cite{Griffiths_2013}:
\begin{eqnarray}
P & = & \frac{\langle S \rangle}{c},
\end{eqnarray}
where $\langle S \rangle$ is the modulus of the time-averaged Poynting vector:
\begin{eqnarray}
\label{eq:Time_averaged_Poynting_vector}
\langle S \rangle & = & \frac{1}{T}\int_{0}^{T} dt\; \frac{1}{\mu_{0}}\left|\tilde{\mathbf{E}}\times \tilde{\mathbf{B}} \right|.
\end{eqnarray}
The time average can be done for any period $T$ over which the fields $\tilde{\mathbf{E}}$ and $\tilde{\mathbf{B}}$ oscillate an integer number of $(2\pi)$.  Substituting Eqs.~(\ref{eq:Solution_E_farfield_simplified_final})-(\ref{eq:Solution_B_farfield_simplified_final}) into Eq.~(\ref{eq:Time_averaged_Poynting_vector}), we finally obtain the radiation pressure at the hilt of saber 1 due to the reflected light from  the interaction region:
\begin{eqnarray}
\label{eq:Radiation_pressure_final}
P & = & \frac{\mu_{0}}{8\pi^{2}}\frac{a^{2}}{c^2}E_{1}^{2}E_{2}^{4}\; \omega_{1}^{4} \left(\frac{b^{4}}{d^{2}}\right) \frac{\sin^{2}\left(k_{1}b\right)}{k_{1}^{2}}.
\end{eqnarray}
The expression for the radiation pressure at the hilt of saber 2 is similar (with indices swapped $1 \leftrightarrow 2$).

\section{Results and discussion}

To get an order-of-magnitude estimate of the radiation pressure at the hilt of saber 1, we assume some typical numbers based on the Star Wars movies and the geometry shown in Fig.~\ref{fig:Crossing_beams_geometry}.  Lightsabers blades are about one meter in length, so we take $d = 0.5$ m.  In addition, we take the diameter of each beam to be $b = 0.02$ m.  We also assume that lightsaber 1 is wielded by Luke Skywalker, implying that its wavelength is approximately $\lambda_{1} = 460$ nm (i.e. in the blue range).  For simplicity, we assume that both lightsabers have equal electric field strengths ($E_{1} = E_{2}$).  Note that the value of $\lambda_{2}$ is not necessary here, because the dominant contribution to the radiation pressure~(\ref{eq:Radiation_pressure_final}) comes from the light of lightsaber 1 reflected back toward the hilt at the interaction region (term $|\mathbf{P}_{1}|$).

\begin{figure}[ht]
\centering
\includegraphics[width=0.55\textwidth]{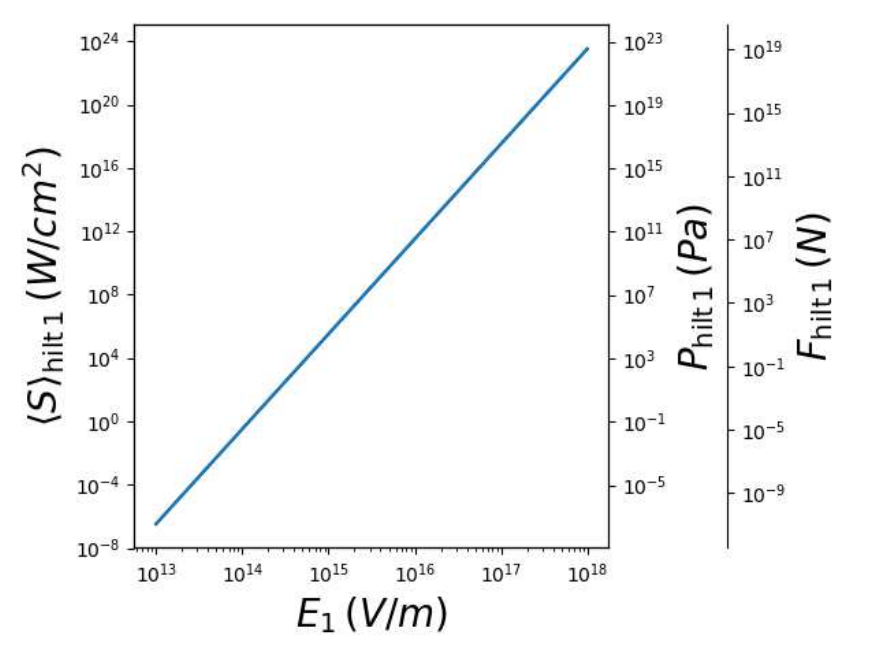}
\caption{Time-averaged Poynting vector, radiation pressure and force at the hilt of saber 1 due to the reflected light coming from saber 2.  We assumed $E_{1} = E_{2}$ and used the following values for the plots: $d = 0.5$ m, $b = 0.02$ m, $\lambda = 460$ nm.}
\label{fig:Pressure_force_at_hilt1}
\end{figure}

Figure~\ref{fig:Pressure_force_at_hilt1} shows the time-averaged Poynting vector and radiation pressure  at the hilt of saber 1 as a function of the field strength, as obtained from Eqs.~(\ref{eq:Time_averaged_Poynting_vector}) and (\ref{eq:Radiation_pressure_final}).  Note that the Euler-Heisenberg effective theory used to perform the calculation (see Sect.~\ref{sec:Nonlinear_em}) is valid for electric fields smaller than the Schwinger field $E_{\rm crit} \approx 1.3 \times 10^{18}$ V/m; we thus limit the field strength to values lower than the Schwinger field in Fig.~\ref{fig:Pressure_force_at_hilt1}.   Note also that as the electric field gets closer to $E_{\rm crit}$, higher order corrections to Eqs.~(\ref{eq:Time_averaged_Poynting_vector}) and (\ref{eq:Radiation_pressure_final}) become larger.  Consequently, our numerical results become less accurate closer to the critical electric field.    The force exerted on the hilt of saber 1 by the reflected light from the interaction region can be obtained by multiplying the radiation pressure with the area of the beam.  The results are shown in Fig.~\ref{fig:Pressure_force_at_hilt1}.  From the above, we estimate that an electric field strength of about $E_{1}\sim 10^{15}$ V/m is required to exert a force of $10$ N on the hilt (roughly equivalent to the force exerted by a one kilogram object falling on your foot)\footnote{Such a large force might seem surprising to specialists in high-energy lasers.  The key difference between the lightsabers described here and real high-energy laser experiments is the size of the interaction region (typically of the order of the laser's wavelength, or $\lambda\sim800$ nm).  Since  radiation pressure depends on the size of the interaction region to the fourth power (see Eq.~(\ref{eq:Radiation_pressure_final})), real high-energy laser experiments would produce pressures that are $(\lambda/b)^{4}\sim 10^{-17}$ times smaller than lightsabers for the same intensities.}.  This corresponds to field strengths at the limit of present-day petawatt lasers.  It is also important to note that existing petawatt lasers deliver ultra intense pulses of extremely short durations (typically 10-100 fs)\cite{danson2015} in very small volumes (the highest fields are obtained by tightly focusing the laser beam to the diffraction limit with a volume of $\sim \lambda^{3}$), while lightsabers require high intensities for sustained periods of time in a large volume.  We thus conclude that, even though not physically impossible, producing an apparent feeling of solidity between two lightsaber blades would require lasers that are far beyond present day capabilities.


\begin{figure}[ht]
\centering
\includegraphics[height=3in]{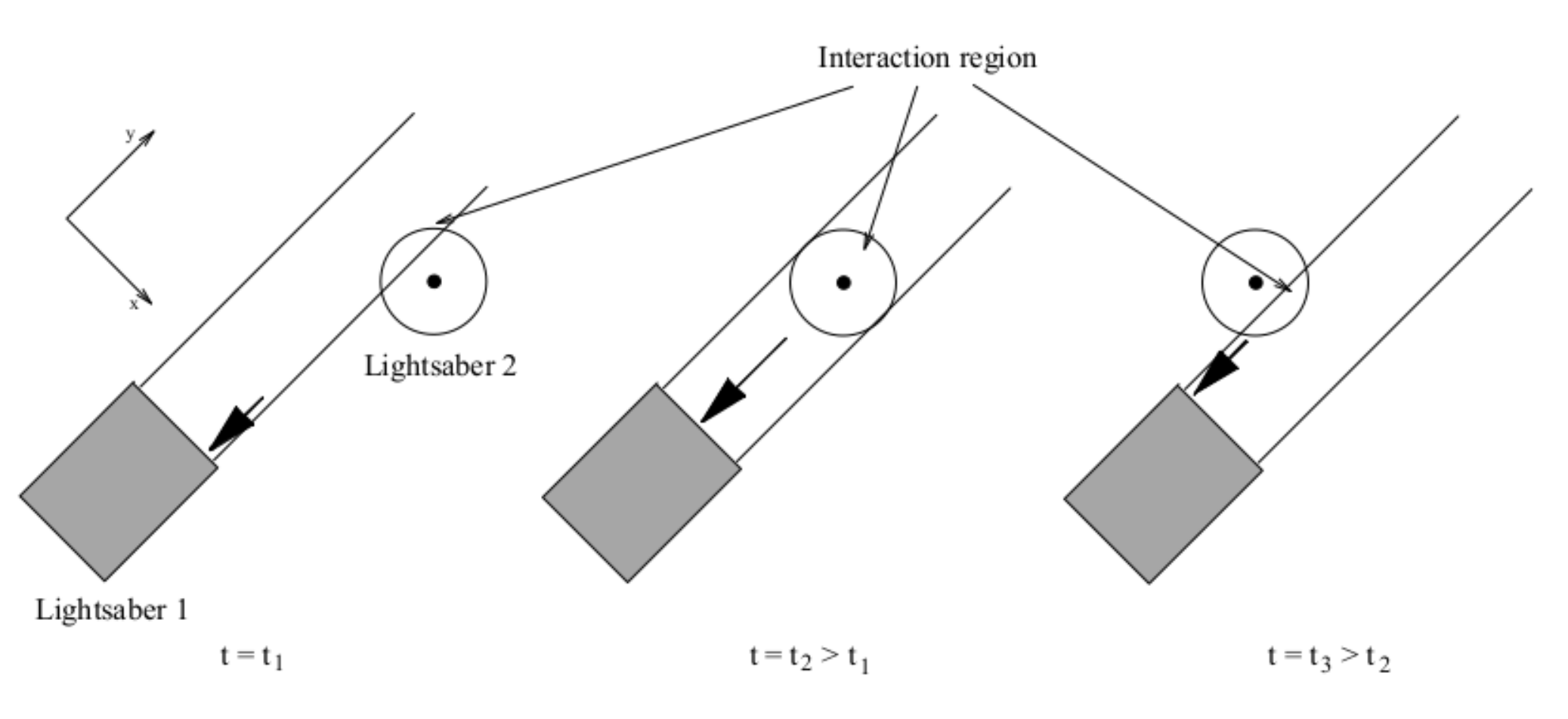}
\caption{Typical clash between two lightsaber beams for three different times ($t_{1} < t_{2} < t_{3}$).  Lightsaber 1 moves to the right, while lightsaber 2 moves to the left.  The thick black arrows indicate where most of the reflected photons coming from the interaction region hit the hilt of lightsaber 1.}
\label{fig:Lightsaber_beam_clash}
\end{figure}

Note also that the sensation felt by the lightsaber wielder would be different from a normal clash between two solid blades.  The argument goes as follows.  Figure~\ref{fig:Lightsaber_beam_clash} shows a typical clash between two lightsaber beams at three different times.  When the two beams start to overlap (Fig.~\ref{fig:Lightsaber_beam_clash}, left panel), the interaction region is small, and the reflected photons coming from the interaction region traveling parallel to the beam of lightsaber 1 only hit a small portion of its hilt\footnote{This is a simplification of the full situation, since photons from the interaction region can be reflected in all directions.}.  This produces a torque on the hilt of lightsaber 1, making it rotate clockwise.  At a later time, the two beams fully overlap (Fig.~\ref{fig:Lightsaber_beam_clash}, center panel), and the reflected photons hit the hilt of lightsaber 1 symmetrically.  This results in a net force in the direction of the reflected photons, with no net torque on the hilt.  Similarly, just before the two beams stop overlapping (Fig.~\ref{fig:Lightsaber_beam_clash}, right panel), the reflected photons traveling parallel to the beam of lightsaber 1 only hit a small portion of its hilt, thus producing a net torque that makes it rotate counterclockwise.  The above sequence of torques and forces felt by the lightsaber wielder is very peculiar.  In comparison, if the two blades in Fig.~\ref{fig:Lightsaber_beam_clash} would be solid, the clash between the two blades would make blade 1 rotate in the counterclockwise direction.   The peculiar sensation felt by the wielder is to be expected, since the interaction between the two light blades does not come from the rigidity of the blades, but from scattered light.  Depending on the geometry (i.e. how the light blades hit each other) and on the intensity of the light, the force felt by the wielder might be able to slow the motion of the blade dramatically.  So even without blade rigidity, it might still feel ``almost like'' a solid blade.

Given the ultrahigh laser intensities discussed above, it is also interesting to study the amount of energy necessary to power such lightsabers.  From the modulus of the time-averaged Poynting vector for a monochromatic plane wave~\cite{Griffiths_2013}:
\begin{eqnarray}
\label{eq:Poyting_vector_plane_wave}
\langle S \rangle_{\rm plane\; wave} & = & \frac{1}{2}c\epsilon_{0} E_{1}^{2},
\end{eqnarray}
we can obtain the power per unit area emitted by a lightsaber as a function of the field intensity.  For $E_{1}\sim 10^{15}$ V/m (sufficient to produce a force of roughly $10$ N on the hilt), we get $\langle S \rangle_{\rm plane\; wave} \sim 1.3 \times 10^{23}$ W/cm$^{2}$.  Multiplying the time-averaged Poynting vector by the cross-sectional area of a beam, we obtain the total power emitted by each lightsaber, or $4\times 10^{23}$ W.  This is three orders of magnitude less than the total luminosity of the Sun, implying that powering such a lightsaber requires a tremendous amount of energy.  For instance, to power one lightsaber for one minute requires $2.4 \times 10^{25}$ J, or one order of magnitude less than the total energy output of the Sun in one second.  

It is clear that an efficient source of energy is needed to power such a lightsaber.  In spite of its great promises, nuclear fusion (i.e. the energy source of the Sun) is not sufficient, since the energy release in a typical fusion reaction (say deuterium-tritium fusion) is $17.6$ MeV per reaction\cite{Krane_1988}, corresponding to roughly $3\times 10^{14}$ J of energy per kilogram of deuterium+tritium.  Thus $8\times 10^{10}$ kg of deuterium+tritium matter is required to power a lightsaber for one minute.  This is a very large mass (about ten times the Great Pyramid of Giza), too large to fit into the hilt of a lightsaber.  A more efficient way of producing energy is matter-antimatter annihilation.  For instance, electron-positron annihilation releases $1.6\times 10^{13}$ J per reaction in the form of high energy photons, or $9\times 10^{16}$ J per kilogram of matter-antimatter.  Assuming that all that energy can be collected at 100\% efficiency, approximately $2.7\times 10^{8}$ kg of matter-antimatter is required to power a lightsaber for one minute.  Since matter-antimatter annihilation is the most efficient way of producing energy per unit mass, $10^{8}$ kg is the minimum mass of ``fuel'' a lightsaber must contain in order to function for a sustained period of time.

The required intensities and energy needs of a lightsaber pose practical problems to their use.  One obvious problem is the weight of the lightsaber, making it hard to lift it, let alone fight with it.  Even more worrisome is the extreme gravitational force exerted by the lightsaber hilt when trying to grasp it.  A person with a mass of $70$ kg would feel a force of $1.2$ N at a distance of $1$ m from the hilt (a gentle tugging), while the same person would feel a force of $12000$ N at $1$ cm from the hilt.  In the colorful language of Randall Munroe's ``What if'' book\cite{Munroe_2014}, ``when your fingertip actually comes into contact with the (hilt), the pressure in your fingertips becomes too strong, and your blood breaks through your skin. [...] your arm remains attached to your body---flesh is surprisingly strong---but blood pours from your fingertip much faster than ordinarily possible.''  This type of extreme behavior is not surprising, since the density of the hilt is $8.6 \times 10^{12}$ kg/m$^{3}$, somewhere between the density of white dwarfs ($10^{9}$ kg/m$^{3}$ ) and neutron stars ($10^{17}$ kg/m$^{3}$).

Another important problem is the recoil of the lightsaber when starting it.  For an electric field intensity of $E_{1}\sim 10^{15}$ V/m, the radiation pressure corresponding to a plane wave is\cite{Griffiths_2013} $P = \langle S\rangle_{\rm plane\;wave}/c \sim 4.3\times 10^{18}$ Pa.  Multiplying the radiation pressure by the cross-sectional area of the beam gives the recoil force of the lightsaber $F_{\rm recoil}\sim 1.4\times 10^{15}$ N.  As a comparison, the thrust of a Saturn V rocket at liftoff is $3.5\times 10^{7}$ N.    Thus due to the high recoil, such a lightsaber would be very difficult to wield (although it could very well be used as a propulsion system for starships).






\section{Conclusion}

In this paper, we have shown that two lightsaber beams can interact and give an apparent feeling of solidity to the wielders, as portrayed in the Star Wars movies.  The interaction between the beams is not direct, but mediated through interaction with virtual particles in the vacuum that add nonlinear source terms to Maxwell's equations.  This type of light-light interaction requires lasers of enormous power, beyond the capabilities of present day technology.  Said differently, we have shown that this aspect of lightsaber behavior is  not impossible due to limitations of the laws of physics, but is very implausible due to the high intensities and energy needed for their operation.  The other fundamental problem of lightsabers (i.e. finiteness of the blade) is still unsolved, but the research on ``needles of light'' mentioned in the Introduction seems a promising avenue to tackle this challenge. 

Even more importantly, we presented a science-fiction example of problem in nonlinear electrodynamics that is sufficiently simple to be solved by undergraduates in physics (either analytically, or using numerical methods).  The method used in this paper is similar to the ones used in calculations of vacuum processes in high intensity lasers, at the cutting edge of present day research.  It shows the power of using science-fiction themes in the classroom, where it is possible to study interesting problems without being bothered by physical plausibility.  We argue that using science-fiction themes opens up a whole new class of examples and problems that have the potential of increasing student engagement.  Moreover, this class of problems can lead to different sets of assumptions and reasonings, which can deepen students' understanding of physics.





\appendix   

\section{Solutions for the generated fields}
\label{app:Solutions_generated_fields}

To find the solutions to Eqs.~(\ref{eq:Wave_equation_1_Egenerated})-(\ref{eq:Wave_equation_2_Bgenerated}), we  first Fourier transform the fields in time:
\begin{eqnarray}
\label{eq:FT_E}
\tilde{\mathbf{E}}(\mathbf{r},t) & = & \int \frac{d\omega}{(2\pi)} e^{-i\omega t}\; \tilde{\mathbf{E}}(\mathbf{r},\omega), \\
\label{eq:FT_B}
\tilde{\mathbf{B}}(\mathbf{r},t) & = & \int \frac{d\omega}{(2\pi)} e^{-i\omega t}\; \tilde{\mathbf{B}}(\mathbf{r},\omega),
\end{eqnarray}
(with similar Fourier transforms for the sources $\mathbf{S}_{1}$ and $\mathbf{S}_{2}$) and substitute expressions~(\ref{eq:FT_E})-(\ref{eq:FT_B}) into Eqs.~(\ref{eq:Wave_equation_1})-(\ref{eq:Wave_equation_2}) to obtain Helmholtz's equations:
\begin{eqnarray}
\left(\nabla^{2} + \frac{\omega^{2}}{c^{2}}\right)\tilde{\mathbf{E}}(\mathbf{r},\omega) & = & -\mathbf{S}_{1}(\mathbf{r},\omega), \\
\left(\nabla^{2} + \frac{\omega^{2}}{c^{2}}\right)\tilde{\mathbf{B}}(\mathbf{r},\omega) & = & -\mathbf{S}_{2}(\mathbf{r},\omega).
\end{eqnarray}
As explained in details in Ref.~\cite{Jackson_1999} (p. 243), we can use Green's functions techniques to solve for the (Fourier transformed) generated fields in terms of the sources.  The result is:
\begin{eqnarray}
\label{eq:Solution_E_FT}
\tilde{\mathbf{E}}(\mathbf{r},\omega) & = & \frac{1}{4\pi}\int_{\mathbb{R}^{3}}d^{3}l\; \frac{e^{\frac{i\omega|\mathbf{r}-\mathbf{l}|}{c}}}{|\mathbf{r}-\mathbf{l}|} \;\mathbf{S}_{1}(\mathbf{l},\omega), \\
\label{eq:Solution_B_FT}
\tilde{\mathbf{B}}(\mathbf{r},\omega) & = & \frac{1}{4\pi}\int_{\mathbb{R}^{3}}d^{3}l\; \frac{e^{\frac{i\omega|\mathbf{r}-\mathbf{l}|}{c}}}{|\mathbf{r}-\mathbf{l}|} \;\mathbf{S}_{2}(\mathbf{l},\omega).
\end{eqnarray}
Fourier transforming back to real time, we obtain:
\begin{eqnarray}
\label{eq:Solution_E}
\tilde{\mathbf{E}}(\mathbf{r},t) & = & \int\frac{d\omega}{(2\pi)} e^{-i\omega t} \left( \frac{1}{4\pi}\int_{\mathbb{R}^{3}}d^{3}l\; \frac{e^{\frac{i\omega|\mathbf{r}-\mathbf{l}|}{c}}}{|\mathbf{r}-\mathbf{l}|} \;\mathbf{S}_{1}(\mathbf{l},\omega)\right), \\
\label{eq:Solution_B}
\tilde{\mathbf{B}}(\mathbf{r},t) & = & \int\frac{d\omega}{(2\pi)} e^{-i\omega t} \left(\frac{1}{4\pi}\int_{\mathbb{R}^{3}}d^{3}l\; \frac{e^{\frac{i\omega|\mathbf{r}-\mathbf{l}|}{c}}}{|\mathbf{r}-\mathbf{l}|} \;\mathbf{S}_{2}(\mathbf{l},\omega)\right).
\end{eqnarray}
with the Fourier transformed sources:
\begin{eqnarray}
\label{eq:Source_term_1_FT}
\mathbf{S}_{1}(\mathbf{r},\omega) & = & \mu_{0}\left(i\omega[\nabla\times\mathbf{M}_{\rm inc}(\mathbf{r},\omega)] + \omega^{2}\mathbf{P}_{\rm inc}(\mathbf{r},\omega) + c^{2}\nabla[\nabla\cdot\mathbf{P}_{\rm inc}(\mathbf{r},\omega)]\right), \\
\label{eq:Source_term_2_FT}
\mathbf{S}_{2}(\mathbf{r},\omega) & = & \mu_{0}\left(-i\omega[\nabla\times\mathbf{P}_{\rm inc}(\mathbf{r},\omega)] - \nabla^{2}\mathbf{M}_{\rm inc}(\mathbf{r},\omega) + \nabla[\nabla\cdot\mathbf{M}_{\rm inc}(\mathbf{r},\omega)] \right)
\end{eqnarray}
Equations~(\ref{eq:Solution_E})-(\ref{eq:Solution_B}) are hard to integrate, so it is worthwhile to find ways to simplify them.  One way is to note that we are interested in the generated fields far away from the interaction region (see assumption A2).  Using the approximation $|\mathbf{r}-\mathbf{l}| \approx |\mathbf{r}| - \hat{\mathbf{r}}\cdot\mathbf{l}$ and Taylor expanding, we can write Eqs.~(\ref{eq:Solution_E})-(\ref{eq:Solution_B}) in the far-field approximation:
\begin{eqnarray}
\label{eq:Solution_E_farfield}
\tilde{\mathbf{E}}(\mathbf{r},t) & = & \frac{1}{4\pi}\int\frac{d\omega}{(2\pi)} \frac{e^{-i\omega \left(t - \frac{|\mathbf{r}|}{c}\right)}}{|\mathbf{r}|}  \int_{\mathbb{R}^{3}}d^{3}l\; e^{-\frac{i\omega \left(\hat{\mathbf{r}}\cdot\mathbf{l}\right)}{c}} \;\mathbf{S}_{1}(\mathbf{l},\omega) + O\left(\frac{b^{2}}{d^{2}}\right), \\
\label{eq:Solution_B_farfield}
\tilde{\mathbf{B}}(\mathbf{r},t) & = & \frac{1}{4\pi}\int\frac{d\omega}{(2\pi)} \frac{e^{-i\omega \left(t - \frac{|\mathbf{r}|}{c}\right)}}{|\mathbf{r}|}  \int_{\mathbb{R}^{3}}d^{3}l\; e^{-\frac{i\omega \left(\hat{\mathbf{r}}\cdot\mathbf{l}\right)}{c}} \;\mathbf{S}_{2}(\mathbf{l},\omega) + O\left(\frac{b^{2}}{d^{2}}\right),
\end{eqnarray}
A further simplification is obtained by noting that incoming electric and magnetic fields vanish at large distances due to the cutoff functions $F_{1}(\mathbf{r})$, $F_{2}(\mathbf{r})$ (see assumption A1).  Using this fact, it is possible to integrate by parts Eqs~(\ref{eq:Solution_E})-(\ref{eq:Solution_B}) and discard the surface terms.  Doing the integration by parts and using various vector identities, we finally obtain (see Appendix~\ref{app:Simplification_far_field} for details):
\begin{eqnarray}
\label{eq:Solution_E_farfield_simplified}
\tilde{\mathbf{E}}(\mathbf{r},t) & = & \frac{\mu_{0}}{4\pi}\int\frac{d\omega}{(2\pi)} \frac{\omega^{2} e^{-i\omega \left(t - \frac{|\mathbf{r}|}{c}\right)}}{|\mathbf{r}|}  \int_{{\cal V}}d^{3}l\; e^{-\frac{i\omega \left(\hat{\mathbf{r}}\cdot\mathbf{l}\right)}{c}} \nonumber \\
                                 &   & \times \left(-\frac{1}{c} \hat{\mathbf{r}}\times \mathbf{M}_{\rm inc}(\mathbf{l},\omega) + \mathbf{P}_{\rm inc}(\mathbf{l},\omega) - \left[\hat{\mathbf{r}}\cdot \mathbf{P}_{\rm inc}(\mathbf{l},\omega)\right]\hat{\mathbf{r}} \right) , \\
\label{eq:Solution_B_farfield_simplified}
\tilde{\mathbf{B}}(\mathbf{r},t) & = & \frac{\mu_{0}}{4\pi}\int\frac{d\omega}{(2\pi)} \frac{\omega^{2} e^{-i\omega \left(t - \frac{|\mathbf{r}|}{c}\right)}}{|\mathbf{r}|}  \int_{{\cal V}}d^{3}l\; e^{-\frac{i\omega \left(\hat{\mathbf{r}}\cdot\mathbf{l}\right)}{c}} \nonumber \\
                                 &   & \times \left(\frac{1}{c} \hat{\mathbf{r}}\times \mathbf{P}_{\rm inc}(\mathbf{l},\omega) + \frac{1}{c^{2}}\mathbf{M}_{\rm inc}(\mathbf{l},\omega) - \frac{1}{c^{2}}\left[\hat{\mathbf{r}}\cdot \mathbf{M}_{\rm inc}(\mathbf{l},\omega)\right]\hat{\mathbf{r}} \right),
\end{eqnarray}
where $\hat{\mathbf{r}} = \mathbf{r}/|\mathbf{r}|$ is a unit vector, $\mathbf{P}_{\rm inc}(\mathbf{l},\omega)$, $\mathbf{M}_{\rm inc}(\mathbf{l},\omega)$ are the temporal Fourier transforms of the polarization and magnetization, and the integration volume ${\cal V}$ is limited to the cubic interaction region (of size $b$) due to the cutoff functions $F_{1}(\mathbf{r})$, $F_{2}(\mathbf{r})$.

Equations~(\ref{eq:Solution_E_farfield_simplified})-(\ref{eq:Solution_B_farfield_simplified}) gives the generated electric and magnetic fields at any point $\mathbf{r}$ far from the interaction region.  The goal of this paper is to compute the radiation pressure at the hilt of one lightsaber due to the reflected photons coming from the interaction region.  For definiteness, let's compute the electric field at the hilt of saber 1 (the magnetic field is done in the similar way).  Substituting the incoming fields~(\ref{eq:E1})-(\ref{eq:B2}) into the polarization and magnetization equations~(\ref{eq:pol})-(\ref{eq:mag}) and taking the temporal Fourier transform, we obtain (note that $\mathbf{E}_{\rm inc}\cdot\mathbf{B}_{\rm inc}=0$ due to assumption A3):
\begin{eqnarray}
\mathbf{P}_{\rm inc}(\mathbf{l},\omega) & = & \mathbf{P}_{1}(\mathbf{l},\omega) + \mathbf{P}_{2}(\mathbf{l},\omega) + \mathbf{P}_{3}(\mathbf{l},\omega) + \mathbf{P}_{4}(\mathbf{l},\omega) + \mathbf{P}_{5}(\mathbf{l},\omega) + \mathbf{P}_{6}(\mathbf{l},\omega), \\
\mathbf{M}_{\rm inc}(\mathbf{l},\omega) & = & \mathbf{M}_{1}(\mathbf{l},\omega) + \mathbf{M}_{2}(\mathbf{l},\omega) + \mathbf{M}_{3}(\mathbf{l},\omega) + \mathbf{M}_{4}(\mathbf{l},\omega) + \mathbf{M}_{5}(\mathbf{l},\omega) + \mathbf{M}_{6}(\mathbf{l},\omega),
\end{eqnarray}
with each of the six different frequency modes given by:
\begin{align}
\label{eq:Pol1}
\mathbf{P}_{1}(\mathbf{r},\omega) &= \frac{a}{2}  E_{1}E_{2}^{2} \;(2\pi) \left[2\delta(\omega-\omega_{1})e^{ik_{1}y} + 2\delta(\omega+\omega_{1})e^{-ik_{1}y}\right] \hat{\mathbf{x}}, \\
\mathbf{P}_{2}(\mathbf{r},\omega) &= \frac{a}{2}  E_{1}^{2}E_{2} \;(2\pi) \left[2\delta(\omega-\omega_{2})e^{ik_{2}z} + 2\delta(\omega+\omega_{2})e^{-ik_{2}z}\right] \hat{\mathbf{x}}, \\
\mathbf{P}_{3}(\mathbf{r},\omega) &= \frac{a}{2}  E_{1}^{2}E_{2} \;(2\pi) \left[\delta(\omega-2\omega_{1}-\omega_{2})e^{i(2k_{1}y+k_{2}z)} + \delta(\omega+2\omega_{1}+\omega_{2})e^{-i(2k_{1}y+k_{2}z)}\right] \hat{\mathbf{x}} , \\
\mathbf{P}_{4}(\mathbf{r},\omega) &= \frac{a}{2}  E_{1}^{2}E_{2} \;(2\pi) \left[\delta(\omega-2\omega_{1}+\omega_{2})e^{i(2k_{1}y-k_{2}z)} + \delta(\omega+2\omega_{1}-\omega_{2})e^{-i(2k_{1}y-k_{2}z)}\right] \hat{\mathbf{x}}, \\
\mathbf{P}_{5}(\mathbf{r},\omega) &= \frac{a}{2}  E_{1}E_{2}^{2} \;(2\pi) \left[\delta(\omega-\omega_{1}-2\omega_{2})e^{i(k_{1}y+2k_{2}z)} + \delta(\omega+\omega_{1}+2\omega_{2})e^{-i(k_{1}y+2k_{2}z)}\right] \hat{\mathbf{x}}, \\
\mathbf{P}_{6}(\mathbf{r},\omega) &= \frac{a}{2}  E_{1}E_{2}^{2} \;(2\pi) \left[\delta(\omega+\omega_{1}-2\omega_{2})e^{i(-k_{1}y+2k_{2}z)} + \delta(\omega-\omega_{1}+2\omega_{2})e^{-i(-k_{1}y+2k_{2}z)}\right] \hat{\mathbf{x}}, 
\end{align}
and: 
\begin{align}
\mathbf{M}_{1}(\mathbf{r},\omega) &= -\frac{ac}{2}  E_{1}E_{2}^{2} \;(2\pi) \left[2\delta(\omega-\omega_{1})e^{ik_{1}y} + 2\delta(\omega+\omega_{1})e^{-ik_{1}y}\right] \hat{\mathbf{y}}, \\
\mathbf{M}_{2}(\mathbf{r},\omega) &= \frac{ac}{2}  E_{1}^{2}E_{2} \;(2\pi) \left[2\delta(\omega-\omega_{2})e^{ik_{2}z} + 2\delta(\omega+\omega_{2})e^{-ik_{2}z}\right] \hat{\mathbf{z}}, \\
\mathbf{M}_{3}(\mathbf{r},\omega) &= \frac{ac}{2}  E_{1}^{2}E_{2} \;(2\pi) \left[\delta(\omega-2\omega_{1}-\omega_{2})e^{i(2k_{1}y+k_{2}z)} + \delta(\omega+2\omega_{1}+\omega_{2})e^{-i(2k_{1}y+k_{2}z)}\right] \hat{\mathbf{z}}, \\
\mathbf{M}_{4}(\mathbf{r},\omega) &= \frac{ac}{2}  E_{1}^{2}E_{2} \;(2\pi) \left[\delta(\omega-2\omega_{1}+\omega_{2})e^{i(2k_{1}y-k_{2}z)} + \delta(\omega+2\omega_{1}-\omega_{2})e^{-i(2k_{1}y-k_{2}z)}\right] \hat{\mathbf{z}}, \\
\mathbf{M}_{5}(\mathbf{r},\omega) &= -\frac{ac}{2}  E_{1}E_{2}^{2} \;(2\pi) \left[\delta(\omega-\omega_{1}-2\omega_{2})e^{i(k_{1}y+2k_{2}z)} + \delta(\omega+\omega_{1}+2\omega_{2})e^{-i(k_{1}y+2k_{2}z)}\right] \hat{\mathbf{y}}, \\
\label{eq:Mag6}
\mathbf{M}_{6}(\mathbf{r},\omega) &= -\frac{ac}{2}  E_{1}E_{2}^{2} \;(2\pi) \left[\delta(\omega+\omega_{1}-2\omega_{2})e^{i(-k_{1}y+2k_{2}z)} + \delta(\omega-\omega_{1}+2\omega_{2})e^{-i(-k_{1}y+2k_{2}z)}\right] \hat{\mathbf{y}}.
\end{align}  
We want the generated electric field at the hilt of saber 1, corresponding to $\mathbf{r} = -d \hat{\mathbf{y}}$ (see Fig.~\ref{fig:Crossing_beams_geometry}).  From the expressions for the polarization and magnetization~(\ref{eq:Pol1})-(\ref{eq:Mag6}), we immediately see that $\hat{\mathbf{r}} \times \mathbf{M}_{\rm inc} = -(|\mathbf{M}_{2}|+|\mathbf{M}_{3}|+|\mathbf{M}_{4}|)\hat{\mathbf{x}}$ and $\hat{\mathbf{r}}\cdot \mathbf{P}_{\rm inc} = 0$.  The electric field thus becomes:
\begin{eqnarray}
\label{eq:Solution_E_farfield_simplified_1}
\tilde{\mathbf{E}}(-d\hat{\mathbf{y}},t) & = & \frac{\mu_{0}}{4\pi}\int\frac{d\omega}{(2\pi)} \frac{\omega^{2} e^{-i\omega \left(t - \frac{d}{c}\right)}}{d}  \int_{-\frac{b}{2}}^{\frac{b}{2}}dx \int_{-\frac{b}{2}}^{\frac{b}{2}}dy \int_{-\frac{b}{2}}^{\frac{b}{2}}dz \; e^{\frac{i\omega y}{c}} \nonumber \\
                                 &   & \times \left(\frac{1}{c}(|\mathbf{M}_{2}(\mathbf{l},\omega)|+|\mathbf{M}_{3}(\mathbf{l},\omega)|+|\mathbf{M}_{4}|(\mathbf{l},\omega)) + |\mathbf{P}_{\rm inc}(\mathbf{l},\omega)| \right)\hat{\mathbf{x}}.
\end{eqnarray}
In the following, we argue that some terms in Eq.~(\ref{eq:Solution_E_farfield_simplified_1}) are numerically much larger than others when the volume integral is performed.  To show this, compare the size of the $\mathbf{P}_{1}$ and $\mathbf{P}_{2}$ terms:
\begin{eqnarray}
\int_{-\frac{b}{2}}^{\frac{b}{2}}dx \int_{-\frac{b}{2}}^{\frac{b}{2}}dy \int_{-\frac{b}{2}}^{\frac{b}{2}}dz \; e^{\frac{i\omega y}{c}} \;|\mathbf{P}_{1}| & \sim & \int_{-\frac{b}{2}}^{\frac{b}{2}}dx \int_{-\frac{b}{2}}^{\frac{b}{2}}dy \int_{-\frac{b}{2}}^{\frac{b}{2}}dz \; e^{\frac{i\omega y}{c}} e^{i k_{1}y} \;\;\sim\;\; \frac{b^{2}}{k_{1}}, \\
\int_{-\frac{b}{2}}^{\frac{b}{2}}dx \int_{-\frac{b}{2}}^{\frac{b}{2}}dy \int_{-\frac{b}{2}}^{\frac{b}{2}}dz \; e^{\frac{i\omega y}{c}} \;|\mathbf{P}_{2}| & \sim & \int_{-\frac{b}{2}}^{\frac{b}{2}}dx \int_{-\frac{b}{2}}^{\frac{b}{2}}dy \int_{-\frac{b}{2}}^{\frac{b}{2}}dz \; e^{\frac{i\omega y}{c}} e^{i k_{2}z} \;\;\sim\;\; \frac{b}{k_{1}k_{2}}.
\end{eqnarray}
Since $\lambda_{i} = (2\pi)/k_{i}$ is the wavelength of the beams and $b \gg \lambda_{i}$, we conclude that the volume integral of the $\mathbf{P}_{1}$ term  is much greater than the one of the $\mathbf{P}_{2}$ term.  A similar analysis shows that all oscillating terms with a space dependence different from $y$ are suppressed by powers of the wavelength, and are thus negligible\footnote{This conclusion holds unless the exponents of the exponentials are exactly zero.  This happens when the wavevectors are chosen to be certain multiple of each others (so-called matching conditions), which is not the case here.}.   Keeping only the dominant terms, the generated electric field can be written as:
\begin{eqnarray}
\label{eq:Solution_E_farfield_simplified_2}
\tilde{\mathbf{E}}(-d\hat{\mathbf{y}},t) & \approx & \frac{\mu_{0}}{4\pi}\int\frac{d\omega}{(2\pi)} \frac{\omega^{2} e^{-i\omega \left(t - \frac{d}{c}\right)}}{d}  \int_{-\frac{b}{2}}^{\frac{b}{2}}dx \int_{-\frac{b}{2}}^{\frac{b}{2}}dy \int_{-\frac{b}{2}}^{\frac{b}{2}}dz \; e^{\frac{i\omega y}{c}}\; |\mathbf{P}_{1}(\mathbf{l},\omega)| \;\hat{\mathbf{x}},
\end{eqnarray}
which can be integrated easily.  The results are shown in Eqs.~\ref{eq:Solution_E_farfield_simplified_final} and \ref{eq:Solution_B_farfield_simplified_final} in the main text.

\section{Simplification of the far field solutions}
\label{app:Simplification_far_field}

In this appendix, we present details on how to obtain Eqs.~(\ref{eq:Solution_E_farfield_simplified})-(\ref{eq:Solution_B_farfield_simplified}) from Eqs.~(\ref{eq:Solution_E_farfield})-(\ref{eq:Solution_B_farfield}). In particular, we consider the following two integrals:
\begin{align}
I_{1} &:= \int_{\mathbb{R}^{3}} \nabla \times \left[ F(\mathbf{s})\mathbf{A}(\mathbf{s}) \right] e^{-i \omega \frac{\hat{R}}{c} \cdot \mathbf{s}} dV, \\
I_{2} &:= \int_{\mathbb{R}^{3}} \nabla [\nabla \cdot \left[ F(\mathbf{s})\mathbf{A}(\mathbf{s}) \right]] e^{-i \omega \frac{\hat{R}}{c} \cdot \mathbf{s}} dV,
\end{align}
where $\mathbf{A}$ can be either the polarization or magnetization and where $F(\mathbf{s})$ is a cutoff function that vanishes at large distance (see assumption A1). Then, we make use of the following identities for the integrands:
\begin{align}
\nabla \times \left[ F(\mathbf{s})\mathbf{A}(\mathbf{s}) \right] e^{-i \omega \frac{\hat{R}}{c} \cdot \mathbf{s}} &= 
\nabla \times \left[ F(\mathbf{s})\mathbf{A}(\mathbf{s})  e^{-i \omega \frac{\hat{R}}{c} \cdot \mathbf{s}} \right] +
i \omega F(\mathbf{s}) \left[ \frac{\hat{R}}{c} \times  \mathbf{A}(\mathbf{s}) \right] e^{-i \omega \frac{\hat{R}}{c} \cdot \mathbf{s}} , \\
\nabla \left[ G(\mathbf{s}) \right] e^{-i \omega \frac{\hat{R}}{c} \cdot \mathbf{s}} &= 
\nabla  \left[ G(\mathbf{s})  e^{-i \omega \frac{\hat{R}}{c} \cdot \mathbf{s}} \right] +
i \omega   \frac{\hat{R}}{c}   G(\mathbf{s})  e^{-i \omega \frac{\hat{R}}{c} \cdot \mathbf{s}},
\end{align}
along with the following vector calculus identities:
\begin{align}
\int_{V}  \nabla \psi(x) dV &= \oint_{\partial V} \psi(x) d\mathbf{S}, \\
\int_{V}  \nabla \times  \mathbf{V}(x) dV &= \oint_{\partial V} \hat{n} \times \mathbf{V}(x) dS,
\end{align}
in order to re-write the integrals as:
\begin{align}
I_{1} &:= 
\oint_{\partial \mathbb{R}^{3}} F(\mathbf{s})\left[ \hat{n} \times  \mathbf{A}(\mathbf{s}) \right] e^{-i \omega \frac{\hat{R}}{c} \cdot \mathbf{s}} dS
+ i \omega \int_{\mathbb{R}^{3}} F(\mathbf{s}) \left[ \frac{\hat{R}}{c} \times  \mathbf{A}(\mathbf{s}) \right] e^{-i \omega \frac{\hat{R}}{c} \cdot \mathbf{s}} dV, \\
I_{2} &:= \oint_{\partial \mathbb{R}^{3}}  \nabla \cdot \left[ F(\mathbf{s})\mathbf{A}(\mathbf{s}) \right] e^{-i \omega \frac{\hat{R}}{c} \cdot \mathbf{s}} d\mathbf{S}
+ i \omega  \frac{\hat{R}}{c} \int_{\mathbb{R}^{3}} \nabla \cdot \left[ F(\mathbf{s})\mathbf{A}(\mathbf{s}) \right] e^{-i \omega \frac{\hat{R}}{c} \cdot \mathbf{s}} dV.
\end{align}
To simplify the second term of $I_{2}$, we use the identity:
\begin{align}
 \nabla \cdot \left[ F(\mathbf{s})\mathbf{A}(\mathbf{s}) \right] e^{-i \omega \frac{\hat{R}}{c} \cdot \mathbf{s}} =
 \nabla \cdot \left[ F(\mathbf{s})\mathbf{A}(\mathbf{s})  e^{-i \omega \frac{\hat{R}}{c} \cdot \mathbf{s}} \right]
 +i \omega F(\mathbf{s}) \left[ \frac{\hat{R}}{c} \cdot \mathbf{A}(\mathbf{s}) \right] e^{-i \omega \frac{\hat{R}}{c} \cdot \mathbf{s}},
\end{align}
and the divergence theorem:
\begin{align}
\int_{V}  \nabla \cdot  \mathbf{V}(x) dV &= \oint_{\partial V}  \mathbf{V}(x) \cdot  d\mathbf{S}.
\end{align}
After these steps, we get:
\begin{align}
I_{1} &:= 
\oint_{\partial \mathbb{R}^{3}} F(\mathbf{s})\left[ \hat{n} \times  \mathbf{A}(\mathbf{s}) \right] e^{-i \omega \frac{\hat{R}}{c} \cdot \mathbf{s}} dS
+ i \omega \int_{\mathbb{R}^{3}} F(\mathbf{s}) \left[ \frac{\hat{R}}{c} \times  \mathbf{A}(\mathbf{s}) \right] e^{-i \omega \frac{\hat{R}}{c} \cdot \mathbf{s}} dV, \\
I_{2} &:= \oint_{\partial \mathbb{R}^{3}}  \nabla \cdot \left[ F(\mathbf{s})\mathbf{A}(\mathbf{s}) \right] e^{-i \omega \frac{\hat{R}}{c} \cdot \mathbf{s}} d\mathbf{S}
+ i \omega  \frac{\hat{R}}{c} \oint_{\partial \mathbb{R}^{3}}  F(\mathbf{s}) \left[ \mathbf{A}(\mathbf{s}) \right] e^{-i \omega \frac{\hat{R}}{c} \cdot \mathbf{s}} \cdot d\mathbf{S} 
\nonumber \\
&
- \omega^{2}  \frac{\hat{R}}{c} \int_{\mathbb{R}^{3}} F(\mathbf{s})\left[ \frac{\hat{R}}{c} \cdot  \mathbf{A}(\mathbf{s}) \right] e^{-i \omega \frac{\hat{R}}{c} \cdot \mathbf{s}} dV.
\end{align}
Assuming that: 
\begin{align}
	\lim_{R \rightarrow \infty} F(\mathbf{s})|_{|\mathbf{s}|=R} &= 0, \\
	\lim_{R \rightarrow \infty} \nabla F(\mathbf{s})|_{|\mathbf{s}| = R} &= 0,
\end{align}
we can neglect the surface terms and we finally obtain:
\begin{align}
	I_{1} &:= 
 i \omega \int_{\mathbb{R}^{3}} F(\mathbf{s}) \left[ \frac{\hat{R}}{c} \times  \mathbf{A}(\mathbf{s}) \right] e^{-i \omega \frac{\hat{R}}{c} \cdot \mathbf{s}} dV, \\
	I_{2} &:= 
	- \omega^{2}  \frac{\hat{R}}{c} \int_{\mathbb{R}^{3}} F(\mathbf{s})\left[ \frac{\hat{R}}{c} \cdot  \mathbf{A}(\mathbf{s}) \right] e^{-i \omega \frac{\hat{R}}{c} \cdot \mathbf{s}} dV.
\end{align}
The above identities allow one to pass directly from Eqs.~(\ref{eq:Solution_E_farfield})-(\ref{eq:Solution_B_farfield}) to Eqs.~(\ref{eq:Solution_E_farfield_simplified})-(\ref{eq:Solution_B_farfield_simplified}).

\begin{acknowledgments}
The authors would like to thank S. Maclean, J. Dumont, C. Lefebvre, D. Gagnon and the other members of the SPACE project for many discussions related to vacuum processes and high intensity lasers. The authors would also like to thank George Lucas for giving us Star Wars.
\end{acknowledgments}

\bibliography{bibliography}



\end{document}